\documentclass[aps,pra,10pt,reprint]{revtex4-1}
\usepackage{amsmath} 
\usepackage{amsfonts}
\usepackage{amssymb}
\usepackage{bm} 
\usepackage{graphicx}
\usepackage[caption=false]{subfig}
\usepackage[colorlinks]{hyperref}

\begin{document}
\title{Models for supercritical motion in a superfluid Fermi liquid}
\author{J. A. Kuorelahti}
\author{S. M. Laine}
\email{sami.laine@oulu.fi}
\author{E. V. Thuneberg}
\affiliation{Nano and molecular systems research unit, University of Oulu, FI-90014, Finland}

\date{\today} % leave empty to show no date 
%
%\begin{abstract}
%We study objects moving in a Fermi superfluid at velocities on the order of the Landau velocity $v_L$. We introduce a boundary condition that describes diffuse reflection of quasiparticles on a scale larger than the coherence length. Applying this boundary condition we calculate the drag force on steadily moving objects of different sizes. For a pointlike particle we find a critical velocity at $v_L$. For a macroscopic cylinder we need to take into account the spatially varying flow field. At low velocities this arises from ideal flow of the superfluid. At higher velocities the flow field is modified by excitations that are created when the flow velocity locally exceeds $v_L$. We investigate multiple limiting cases. In the absence of quasiparticle-quasiparticle collisions we find a critical velocity slightly higher than $v_L$. The drag force at $2v_L$ is reduced by an order of magnitude compared to the pointlike particle. In the collision-dominated limit the flow shows signs of instability at a velocity below $v_L$.
%\end{abstract}

\begin{abstract}
We study the drag force on objects moving in a Fermi superfluid at velocities on the order of the Landau velocity $v_L$. The expectation has been that $v_L$ is the critical velocity beyond which the drag force starts to increase towards its normal-state value. This expectation is challenged by a recent experiment measuring the heat generated by a uniformly moving wire immersed in superfluid $^3$He. We introduce the basis for the calculation of the drag force on a macroscopic object using the  Fermi-liquid theory of superfluidity. As a technical tool in the calculations we propose a boundary condition that describes diffuse reflection of quasiparticles from a surface on a scale that is larger than the superfluid coherence length. We calculate the drag force on steadily moving objects of different sizes. For an object that is small compared to the coherence length, we find a drag force that is in accordance with the expectation. For a macroscopic object we need to take into account the spatially varying flow field around the object. At low velocities this arises from ideal flow of the superfluid. At higher velocities the flow field is modified by excitations that are created when the flow velocity locally exceeds $v_L$. The flow field causes Andreev reflection of quasiparticles and thus leads to change in the drag force. We calculate multiple limiting cases for a cylinder-shaped object. In the absence of quasiparticle-quasiparticle collisions we find that the critical velocity is larger than $v_L$ and the drag force (per cross-sectional area) at $2v_L$ is reduced by an order of magnitude compared to the case of a small object. In a collision-dominated limit the flow shows signs of instability at a velocity below $v_L$.
\end{abstract}

\maketitle

\section{Introduction}

For many purposes it would be beneficial to travel fast. The problem is that higher velocities generally  require more power. Often the force needed to move an object increases rapidly beyond a critical velocity. For example, an airplane exceeding the velocity of sound requires more power as it starts to emit a cone of sound waves \cite{Mach1887}. These waves are stationary in the frame fixed to the airplane. Related energy loss mechanism appears for fast charged particles in a medium, observed as Cherenkov radiation, when the particle velocity exceeds the velocity of light in the medium \cite{LL:EDCM}. Similar situation occurs in media supporting waves or elementary excitations which have nontrivial dispersion. For example, there is a critical velocity for emission of  waves on the surface of liquid \cite{Thomson1891,Whitham}. For ships this critical velocity is impractically low, but nevertheless the power consumption of a ship increases rapidly when the ship velocity exceeds the phase velocity of relevant surfaces waves (leading to the concept of hull speed). The leading waves, and the only ones in linear approximation, have a phase velocity whose component in the direction of the ship equals the ship velocity, and thus are stationary in the frame of the ship. The same concept applies to objects moving in superfluids. In this context the condition is known as Landau criterion. Lev Landau suggested that superfluidity results from the absence of these type of stationary elementary excitations \cite{Landau41}. Such a linear-response critical velocity has been observed in the boson superfluid $^4$He under pressure with ions \cite{Allum77}.  There is evidence of a  similar critical velocity in  the fermion superfluid $^3$He, also obtained with ions \cite{Ahonen76}. 

Often the critical velocity derived from a linear theory is considered as an upper limit of low-dissipation motion. Namely, there can be other, more complicated and nonlinear effects that lead to large dissipation already at lower velocities. In particular, there can be eddies in the fluid. In superfluids, the eddies consist of quantized vortices. For most experiments in superfluids with either a moving macroscopic object or flow, the motion becomes dissipative at such low velocities that even achieving the Landau velocity becomes difficult \cite{Donnelly91,Castelijns86,Ruutu97,Onofrio00,Varoquaux15}.

Against this background, it came as a surprise that Bradley et al.\ \cite{Bradley16} reported observation of low-dissipation motion of a macroscopic object in superfluid $^3$He at velocities exceeding twice the Landau critical velocity. The experiment was made in superfluid $^3$He-B at temperatures $T$ well below the superfluid transition temperature $T_c$. The moving object was a wire able to move at a constant velocity for a time span of $\sim 100$ ms. The dissipation increased gradually with increasing velocity, but there were no particular features that could be interpreted as a critical velocity.

The purpose of this communication is to present some theoretical models that are related to fast motion in a Fermi superfluid. We start by setting up the problem and explaining some basic concepts that are used in the calculations (Sec.\ \ref{s.basics}). For a pointlike impurity we calculate the force which, according to expectation, vanishes at $T=0$ for velocities below the Landau velocity, $v< v_L$, and starts to increase toward its normal state value for $v>v_L$ (Sec.\ \ref{s.ion}). The situation is more complicated for an object larger than the coherence-length scale. Firstly, the superfluid flow around the object causes Andreev reflection of quasiparticles. This reduces the number of scattered quasiparticles that are able to escape from the object and thus reduces the force on the object. We demonstrate this by calculating the drag force on a cylinder assuming an ideal-fluid flow field (Sec.\ \ref{s.hydrof}). Secondly, the flow field is modified by local pair breaking already at object velocities less than $v_L$. We calculate self-consistently the velocity field in two extreme cases. In the limit of no collisions between quasiparticles we show that a local supercritical flow could be stable in some range of object velocities (Sec.\ \ref{s.noc}). The opposite extreme is that full equilibrium is achieved through collisions in the region near the object. In this case we find indication of instability, possibly towards vortex formation  (Sec.\ \ref{s.enr}). As a technical tool in the calculations we propose a mesoscopic diffuse-scattering boundary condition in the superfluid state  (Sec.\ \ref{s.bc}).

\section{Basic concepts}\label{s.basics}

\begin{figure}[tb] %  figure placement: top, bottom, (not here or page) 
\centering
\includegraphics[width=0.8\linewidth]{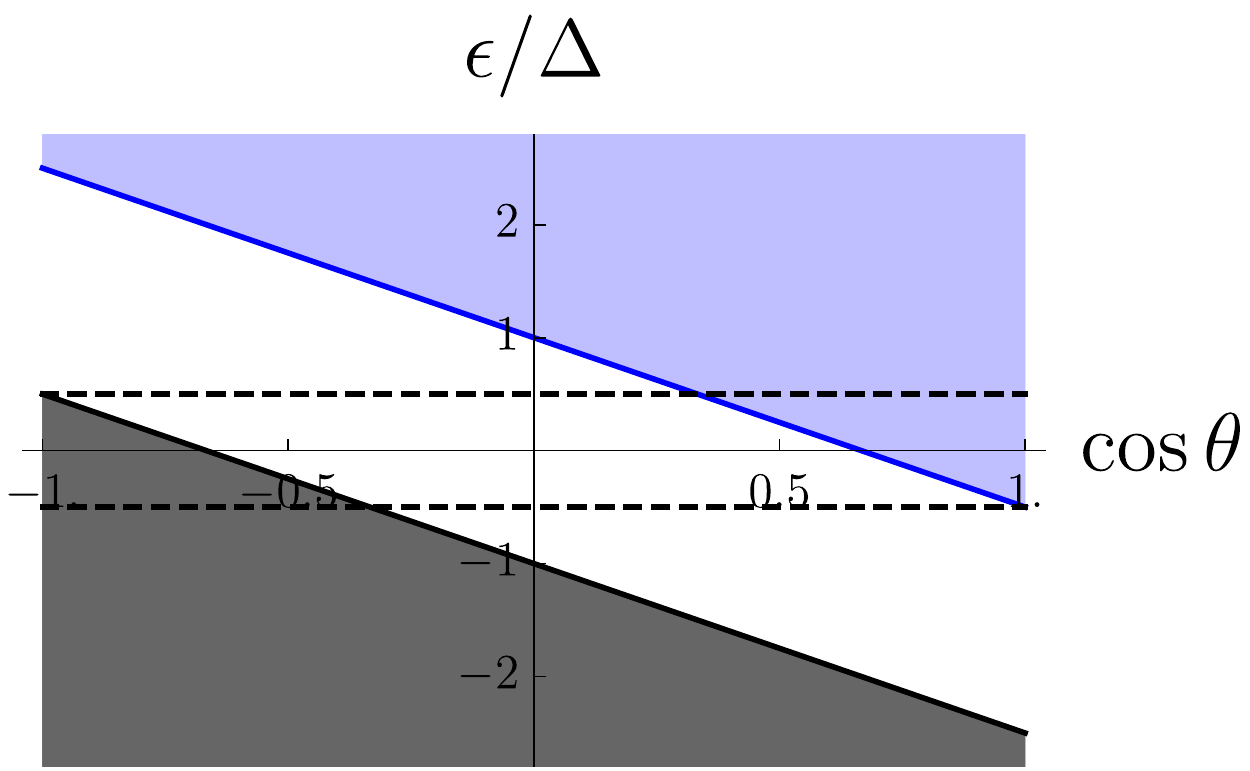}
\caption{The quasiparticle energies as a function of direction cosine $\cos \theta = \hat{\bm p} \cdot \hat{\bm v}$ in a frame moving at  supercritical velocity $v=1.5 v_L$ with respect to the superfluid. A stationary object in this frame can scatter ground state quasiparticles (gray) to the empty excited states (blue) in the energy range limited by the dashed horizontal lines. }
\label{f.BCSExcitationSpectrumC}
\end{figure}

We study a rigid object moving at velocity $v$ in an otherwise stationary medium. By Galilean invariance, this problem is equivalent to a flow of the medium past a stationary object. We call the latter frame of reference the {\em object frame}, which will be very useful in the following. There is an important difference between the case of a single object, the case studied here, and the case of there being a distribution of objects that fill the volume under study. The latter case appears, for example, in impure superconductors under superflow. We briefly return to this topic after discussion of Fig.\ \ref{f.BCSExcitationSpectrumC}.

 In general, there will be a drag force opposing the motion of the object. For example, consider a fluid that can be described by hydrodynamic theory. In the Navier-Stokes equations, the force is caused by viscous terms \cite{LLf,Batchelor}. At low velocities the force is linear in velocity. With increasing velocity the viscous terms and nonlinear convective terms in the Navier-Stokes equations cause the flow to separate from the object surface. This leads to eddies and a wake, which increase the drag. These effects, however, are not our main interest in the following, and thus we assume that they are small. Neglecting both the viscous and the nonlinear terms in the Navier-Stokes equations means that the system is sufficiently described by the linearized Euler equation together with the linearized continuity equation and boundary conditions. Under these assumptions, the force vanishes for an object moving at a constant, low velocity. Below we implicitly make the same assumptions in the more detailed theories we use.

The linearized equations allow also the determination of the elementary excitations of the system. For example, the hydrodynamic equations for a simple fluid have one type of elementary excitation, the longitudinal sound wave. The angular  frequency $\omega$ of the wave is related to the wave vector $\bm k$ by a linear dispersion relation $\omega=ck$, where $c$ is the sound velocity. Let us consider the problem of emission of sound from a rigid object moving at constant velocity $v$. For this we change to the object frame. In this frame the frequency of the excitation is  $\omega'=\omega-\bm v\cdot\bm k$ according to the Galilei transformation. Since the source is stationary in the object frame, the only wave that the object generates in the linear approximation corresponds to zero frequency, $\omega'=\omega-\bm v\cdot\bm k=0$, that is,
\begin{eqnarray}
\omega=\bm v\cdot\bm k.
\label{e.radcond}\end{eqnarray}
This has to be satisfied simultaneously with the dispersion relation $\omega=ck$. As a result, no waves are generated at $v<c$. When $v>c$ there is a wavefront that forms a cone of angle $\alpha$ with $\bm v=v\hat{\bm x}$, so that $\sin\alpha=k_x/k=c/v$. The waves carry off energy, and thus a dissipative force is exerted on the object when $v$ exceeds the critical velocity $c$.

More generally, we can assume a medium with a general dispersion relation $\omega(\bm k)$ of the waves. In quantum mechanics we can alternatively speak of the energy $\epsilon=\hbar \omega$ and momentum $\bm p=\hbar \bm k$ of an elementary excitation. It is standard to define phase velocity $\bm v_p$, the velocity of the wave crests, and group velocity $\bm v_g$, the velocity of a wave packet. For an excitation with wave vector $\bm k$ these are
\begin{eqnarray}
\bm v_p(\bm k)=\frac{\omega(\bm k)}{k}\hat{\bm k},\quad \bm v_g(\bm k)=\bm\nabla_{\bm k}\,\omega(\bm k).
\end{eqnarray}
The condition for a rigid object to emit elementary excitations is the same as was discussed above and given in Eq.\ (\ref{e.radcond}). Thus the emitted waves need to satisfy simultaneously $\omega=\bm v\cdot\bm k$ and $\omega(\bm k)$. As $\bm v\cdot\bm k\le vk$, we must have $v\ge\omega(\bm k)/k$ in order to be able to create an excitation with wave vector $\bm k$. The minimum of the right-hand side is the velocity below which no excitations can be created,
\begin{eqnarray}
v_L=\left(\frac{\omega(\bm k)}{k}\right)_{\rm min}.
\label{e.landauvg}\end{eqnarray}
This is known as the Landau velocity and the condition $v<v_L$ as the Landau criterion for superfluidity \cite{Landau41}.
In order to find the minimum, the gradient of $\omega(\bm k)/k$ with respect of $\bm k$ should vanish. This means that for the excitation created just at the Landau velocity, the phase and group velocities are equal. At a slightly larger $v$ there will be two types of emitted waves, one with group velocity larger than $v_L$ and one with a smaller group velocity.  In order to be able to neglect the viscous force at $v< v_L$, we have to assume that no elementary excitations are excited initially. This means a temperature that is small compared to the minimum of $\hbar\omega$. For simplicity, we take the limit of zero temperature, $T=0$.

The most commonly observed dispersive waves are the waves on the surface of water \cite{Thomson1891,Whitham}. The factors determining the dispersion are gravity and surface tension. These waves have a critical velocity of 23 cm/s (under standard conditions). The wave pattern generated by a ship can to a large extent be explained by linear gravity waves, which have $v_g=v_p/2$.

We now concentrate on a Fermi superfluid. Serene and Rainer have formulated a general quasiclassical approach \cite{Serene83}. For consistency, our notation below is close to theirs. 
The energy of a quasiparticle excitation with momentum $\bm p$ is 
\begin{eqnarray}
\epsilon(\bm p)=\sqrt{[\xi_p+u(\hat{\bm p})]^2+\Delta^2}+a(\hat{\bm p}),
\label{e.edrel}\end{eqnarray}
where $\xi_p=v_F(p-p_F)$, $p_F$ is the Fermi momentum, $v_F=p_F/m^*$ the Fermi velocity, $m^*$ the effective mass, and $\hat{\bm p}=\bm p/p$ the direction of the momentum. The superfluid has energy gap $\Delta$. Compared to Ref.\ \cite{Serene83} we simplify a bit by dropping the $\hat{\bm p}$ dependence of $\Delta$. This can be justified in the B phase of superfluid $^3$He at low velocities. The dispersion relation (\ref{e.edrel}) also depends on quasiparticle potentials $u$ and $a$ (denoted by $\tilde u$ and $\tilde a$ in Ref.\ \cite{Serene83}), which by definition are even and odd in the momentum direction, $u(-\hat{\bm p})= u(\hat{\bm p})$ and $a(-\hat{\bm p})=- a(\hat{\bm p})$. According to the quasiclassical assumption all energies in the theory are small compared to the Fermi energy $\sim v_Fp_F$. That is, $\xi_p$, $\Delta$, $u$, $a$, $\epsilon \ll v_Fp_F$, and we calculate everything only in the leading order of these small quantities. 

In equilibrium in the rest frame of the superfluid, the quasiparticle potentials vanish, $u= a=0$. We see that in this case the minimum excitation energy is $\Delta$, and it is achieved at the Fermi surface $p=p_F$. There are two types of excitations, particle-type with $p>p_F$ and hole-type with $p<p_F$. Their group velocities are $v_F\xi_p\hat{\bm p}/\sqrt{\xi_p^2+\Delta^2}$, which  for hole-type excitations is in the direction opposite to momentum. The Landau criterion (\ref{e.landauvg}) gives the critical velocity
\begin{eqnarray}
v_L=\frac{\Delta}{p_F}
\label{e.landauvf}\end{eqnarray}
with a vanishingly small correction in the quasiclassical approximation \cite{Combescot06}.

\begin{figure}[tb] %  figure placement: top, bottom, (not here or page)
\centering
\subfloat[]{\label{f.BCSExcitationSpectrumA} \includegraphics[width=0.8\linewidth]{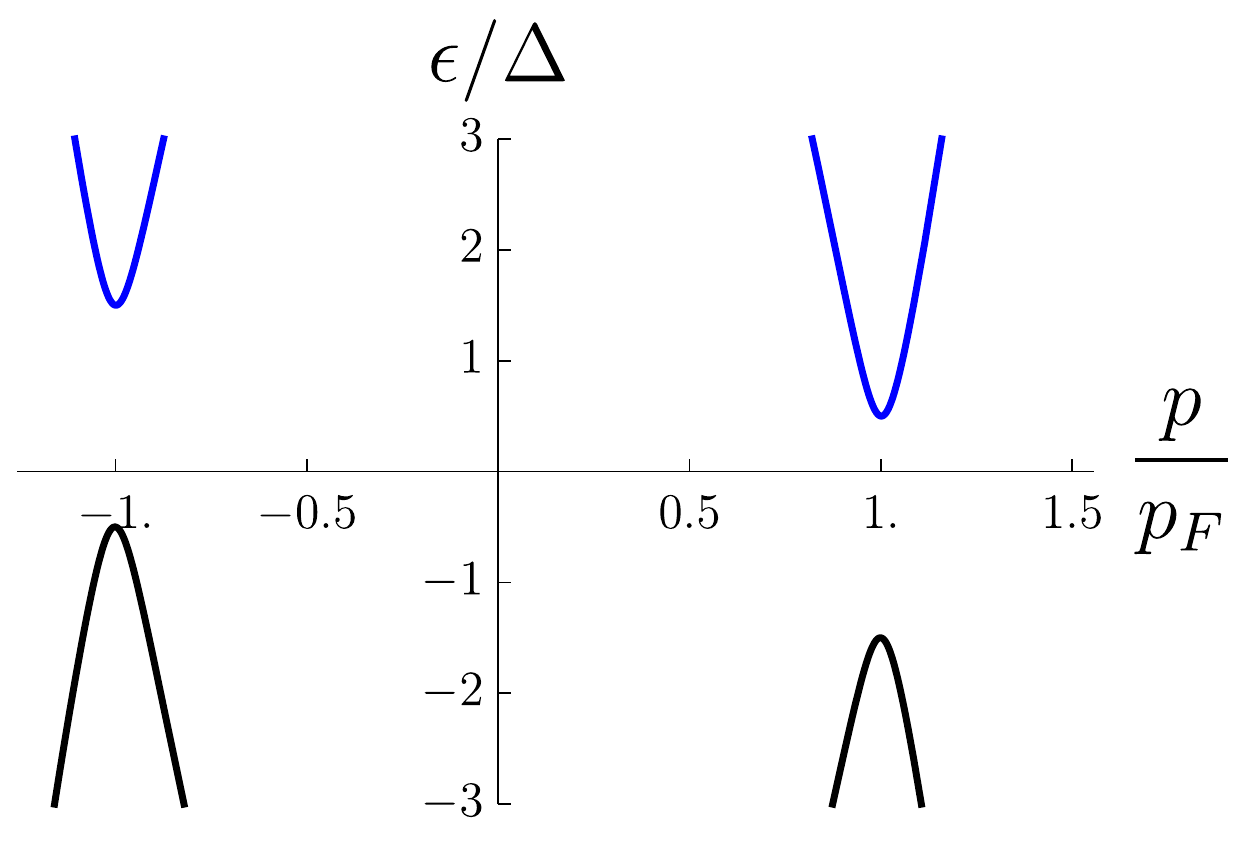}}

\subfloat[]{\label{f.BCSExcitationSpectrumB} \includegraphics[width=0.8\linewidth]{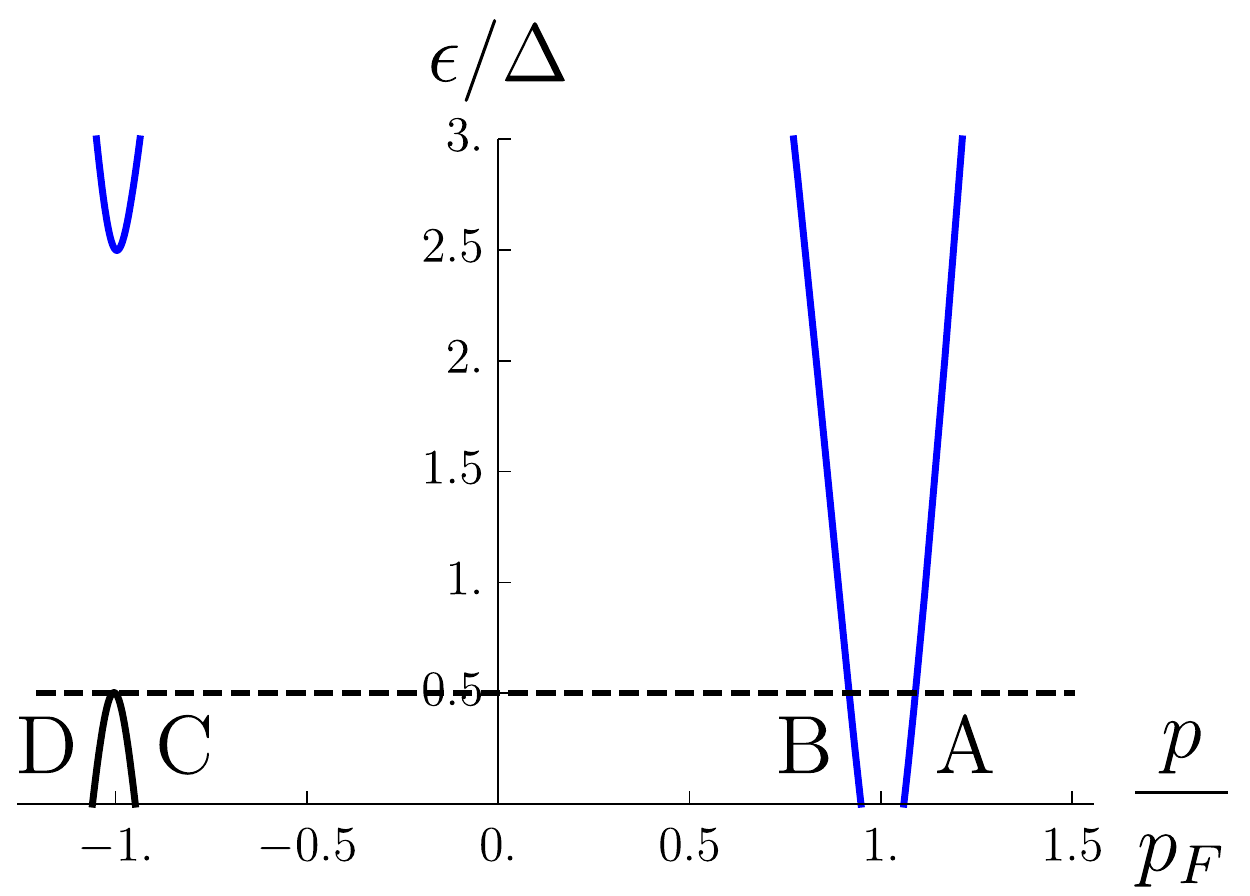}}
\caption{A sketch of the equilibrium excitation spectrum (\ref{e.edrel}) seen from a frame moving at (a) a subcritical velocity $v_k=0.5v_L$ and (b) a supercritical velocity $v_k=1.5v_L$. The abscissa gives the momentum $p$ parallel to an arbitrary direction $\hat{\bm k}$, and $v_k$ is the component of the velocity in that direction. Figure (a) uses the semiconductor picture, which includes also negative-energy states. It has the advantage of showing the filled states (black lines) that lead to supercurrent to the left. Figure (b) uses excitation picture where only positive-energy states are shown. An object at rest in (b) frame leads to scattering from the filled states C and D to the empty states A and B. States A and C are particle-type (motion in the same direction as the momentum), while B and D are hole-type (motion in the direction opposite to the momentum).}
\label{f.BCSExcitationSpectrum}
\end{figure}

It is useful to look at the dispersion (\ref{e.edrel}) in a frame moving at velocity $\bm v$ with  respect to the rest frame of the fluid. In this case $u=0$ but $a=-p_F\hat{\bm p}\cdot\bm v$ [in accordance with the Galilei transformation discussed above in connection with Eq.\ (\ref{e.radcond})]. The dispersion relation at a subcritical velocity is depicted in Fig.\ \ref{f.BCSExcitationSpectrum}(a). The excitation energies are given by the blue lines, which have $\epsilon>0$. In addition, Fig.\ \ref{f.BCSExcitationSpectrum}(a) has black lines corresponding to the negative branch of the square root in (\ref{e.edrel}). These can be considered as quasiparticle states that are filled in the ground state. Removing a fermion from such a state of momentum $\bm p$ is equivalent to an excitation with momentum $-\bm p$.  Considering both the negative and positive energy states is called the {\em semiconductor picture} \cite{Tinkham}. One advantage of this picture is that we can see in Fig.\ \ref{f.BCSExcitationSpectrum}(a) that there are more filled states with negative $p$ than with positive $p$. This just corresponds to superflow with velocity $\bm v_s=-\bm v$, since we are in a frame that is moving at velocity $\bm v$ with respect to the superfluid rest frame.

Figure \ref{f.BCSExcitationSpectrum}(b) describes the dispersion of superfluid quasiparticles seen in a frame moving at a supercritical velocity. In this case the part of the dispersion relation with negative square root (\ref{e.edrel}) has positive energy (branches C and D). These states are filled in superfluid equilibrium state, whereas the states with positive square root are empty. In order to avoid double representation of states, Fig.\ \ref{f.BCSExcitationSpectrum}(b) uses the {\em excitation picture}, where only positive energy states are shown. 

Suppose now that the moving frame is the object frame. The object scatters quasiparticles between  states at the same energy. At a subcritical velocity this has no effect since at a given energy all states are either filled or empty. At a supercritical velocity there can be scattering from the filled states C and D to the empty states A and B. Such scattering causes a drag force on the object. This scattering process is called {\em pair breaking} as it reduces the number of Cooper pairs. It should be noted, however, that this scattering process remains also in the limit $v\gg v_L$ and also in the case $\Delta= 0$. In the latter case there are no pairs and the scattering is just the same as the elastic impurity scattering that causes the electrical resistivity of normal state metals at $T=0$.

Figure \ref{f.BCSExcitationSpectrumC} shows the allowed quasiparticle states for arbitrary momentum direction $\cos \theta = \hat{\bm p}\cdot\hat{\bm v}$. We see that for energies $0<\epsilon<p_Fv-\Delta$ there are incoming ground state quasiparticles in momentum directions $\cos\theta<-(\Delta+\epsilon)/p_Fv$. These are scattered elastically to empty states with momentum directions $\cos\theta>(\Delta-\epsilon)/p_Fv$. The particle-type ground state quasiparticles [C in Fig.\ \ref{f.BCSExcitationSpectrum}(b)] come from the front direction. The hole-type ground state quasiparticles [D in Fig.\ \ref{f.BCSExcitationSpectrum}(b)] come from the back direction, as their propagation direction is opposite to the momentum direction. The outgoing particle-type excitations [A in Fig.\ \ref{f.BCSExcitationSpectrum}(b)] come out predominantly (depending on $\epsilon$ and $v/v_L$) from the front direction and hole-type excitations [B in Fig.\ \ref{f.BCSExcitationSpectrum}(b)] from the back direction. 

Let us comment on the relation of the present problem of a moving impurity with {\em gapless superconductivity}. The latter can appear, for example, in superconductors under superflow \cite{Bardeen62,Tinkham,Zagoskin}. The quasiparticle energies are similar to those presented in Figs.\ \ref{f.BCSExcitationSpectrum} and \ref{f.BCSExcitationSpectrumC}, but the difference is that quasiparticle equilibrium has already been achieved by scattering. That is, all states with positive energies are empty, while those with negative energies are filled. Superfluid can still flow in this state as long as some pairs survive, keeping the quasiparticle energies asymmetric. Applied to a p-wave superfluid, namely $^3$He-B, this problem has been studied in Refs.\ \cite{Vollhardt80,Kleinert80}. These studies imply that superfluid flow is in principle possible at velocities exceeding the Landau velocity (\ref{e.landauvf}), but they do not imply an increased critical velocity for a moving object\cite{Baym12}. Another case of gapless superconductivity appears in systems with gap nodes, for example, in the A phase of superfluid $^3$He \cite{Leggett75}. There superflow is possible although $v_L$ vanishes. We are not aware of calculations of the pair breaking by moving objects in such systems.
 
In Sec. \ref{s.bc} we formulate a simple boundary condition that can be used to calculate the distribution of scattered quasiparticles. In Sec.\ \ref{s.ion} this is used to calculate the force on a small object. By small we mean in comparison with the superfluid coherence length $\xi_0=\hbar v_F/2\pi k_BT_c$.

\begin{figure}[tb] %  figure placement: top, bottom, (not here or page)
\centering
\includegraphics[width=0.8\linewidth]{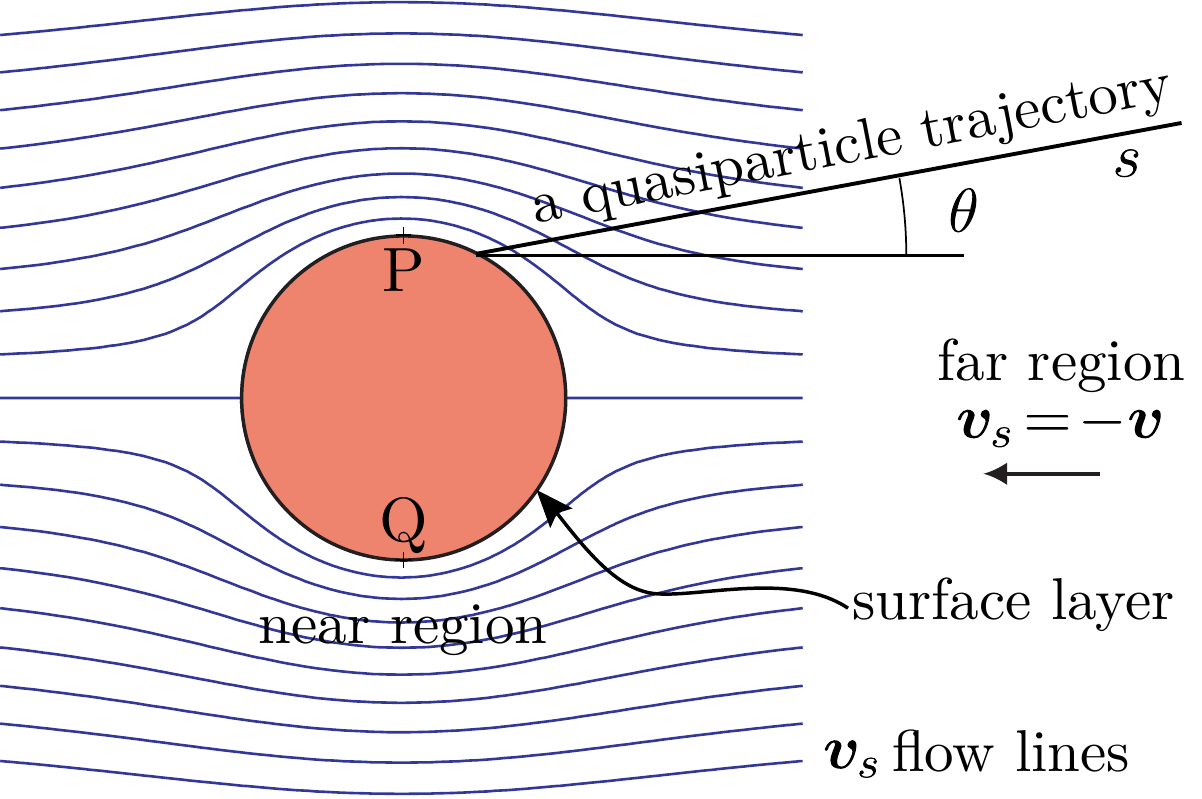} 
\caption{A sketch of a wire moving in a superfluid. The curves are flow lines of the superfluid seen in the rest frame of the wire according to the ideal fluid model (\ref{e.iflff}), see also Fig. \ref{fig:alphafield}(a). The maximal velocity is reached at points P and Q where the velocity $\bm v_s(\bm r)$ is twice as large as far from the wire. The near and far regions and a thin surface layer on the wire surface are indicated. An example quasiparticle trajectory hitting the wire is shown. }
\label{f.CylinderFlow}
\end{figure}

The moving object we consider in particular (besides the small one) is a circular cylinder of radius $R$ to mimic a wire. We assume that the cylinder diameter $2R$ is large in comparison with the superfluid coherence length, $R\gg\xi_0$. We assume the cylinder is moving at velocity $v$ perpendicular to its axis in an initially stationary superfluid. We need to define three different regions, which are illustrated in Fig.\ \ref{f.CylinderFlow}: 1) {\em The surface layer} at the cylinder surface of thickness on the order of $\xi_0$.  2) {\em The near region} around the cylinder of size on the order of the cylinder radius $R$. 3) {\em The far region} at distances $r\gg R$. 

The surface layer has bound quasiparticle states at energies below the gap $\Delta$. The bound quasiparticles collide with the wire wall at intervals $\xi_0/v_F\sim 10^{-9}$ s. This time is short compared to the oscillation or acceleration time scales of the wire, which are $10^{-3}$ s or longer. Thus we assume that the distribution of these states always remains in equilibrium in the cylinder frame. It should be noted that this assumption automatically excludes the dissipation mechanism proposed by Lambert \cite{Lambert90,Lambert92} and similar arguments presented in later work \cite{Volovik09,Bradley16,Zheng17}. Another objection against this mechanism is that the bound states are not found to cross zero energy, at least for small objects \cite{Ashauer88}.

The flow around the cylinder has to satisfy mass conservation. In time-independent case this means that the divergence of the mass current vanishes,
\begin{eqnarray}
\bm\nabla\cdot\bm j=0.
\label{e.ndjeqo}
\end{eqnarray}
At $T=0$ and $v_s<v_L$, all the flow is superflow, $\bm j=\bm j_s=\rho_s\bm v_s$, and the superfluid density equals the liquid density, $\rho_s=\rho=mn_f$. Here $m$ is the fermion mass and $n_f=p_F^3/3\pi^2\hbar^3$ their equilibrium number density. The superfluid velocity is given by the gradient of the phase,
\begin{eqnarray}
\bm v_s=\frac{\hbar}{2m}\bm\nabla\psi.
\label{e.vsgp}\end{eqnarray}
These imply the Laplace equation $\nabla^2\psi=0$, and one easily finds the flow field around the cylinder,
\begin{eqnarray}
\frac{\hbar}{2m}\psi(\bm r)=-v \cos\varphi\left(\frac{R^2}{r}+r\right).
\label{e.iflff}
\end{eqnarray}
This gives $\bm v_s$ in the cylinder frame. We have represented $\bm r$ in cylindrical coordinates $(r,\varphi,z)$ with the cylinder at the origin aligned along the $z$-axis. Some flow lines are drawn in Fig.\ \ref{f.CylinderFlow}. It is noteworthy that the maximum velocity, which is reached at points P and Q, is twice the velocity $v$ in the far region. 
 
\begin{figure}[tb] 
\centering
\includegraphics[width=0.95\linewidth]{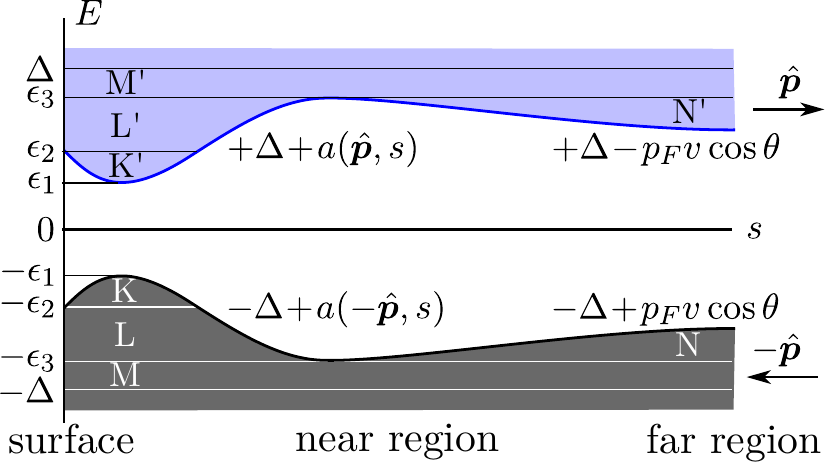} 
\caption{The allowed quasiparticle energies on a trajectory. Here $s$ is the parameter along the trajectory measured from the point on the object surface (Fig.\ \ref{f.CylinderFlow}). The figure gives  filled quasiparticle states with momentum to the left that are lifted from the lower gap edge $E=-\Delta$, as in Fig.\ \ref{f.BCSExcitationSpectrum}(a). These are denoted by K, L, M and N. Correspondingly, there are empty quasiparticle states with momentum to the right whose energies are reduced from the upper gap edge $E=+\Delta$, denoted by K$'$, L$'$, M$'$ and N$'$. The filled states are responsible for the superflow past the object. The figure corresponds to a low velocity where $\epsilon_1$, $\epsilon_2$ and $\epsilon_3$ are all positive. With increasing velocity the energies $\epsilon_i$ decrease and cross zero. With $\epsilon_1<0$ the filled states K are lifted above the empty states K$'$ but there is no scattering between these states as they are not in contact with the object surface (as long as we neglect quasiparticle-quasiparticle scattering). With $\epsilon_2<0$ the particles at states L start to scatter to states L$'$ (and corresponding states on other trajectories hitting the same point on the surface). The reduced current carrying capacity of states L leads to redistribution of current pattern in the near region of the object.  With $\epsilon_3<0$, the scattered quasiparticles can escape from the near region to the far region, and lead to dissipative force on the object.  The velocity required for $\epsilon_3$ to cross zero is higher than the Landau velocity $v_L=\Delta/p_F$.  }
\label{f.trajectoryenergy} 
\end{figure}

The flow field around a macroscopic object causes the quasiparticle energy (\ref{e.edrel}) to vary locally. A possible variation of the allowed energies on a quasiparticle trajectory is depicted in Fig.\ \ref{f.trajectoryenergy}. Consider first a thermally excited hole-type quasiparticle entering from the right. In the energy range M$'$ it hits the wire and is scattered there. In the energy range N$'$ it cannot reach the wire since at some locations in the near region of the wire its energy is not in the allowed range. Instead, the quasiparticle will be Andreev reflected back as a particle-type quasiparticle \cite{Andreev64}. There is only a small momentum transfer in Andreev reflection, $|\bm p'-\bm  p|\ll p_F$, and the momentum is transferred to the superfluid condensate, not the wire. Thus, when an incoming quasiparticle is Andreev reflected before reaching the wire surface, it does not contribute to the force on the wire. This effect has been extensively studied by the Lancaster group in order to calculate the thermal damping force on a wire \cite{Fisher89,Fisher91,Enrico95,Enrico96}. We point out that Andreev reflection is important also at $T=0$ at large velocities. Namely, at the Landau velocity the  states N$'$ start shifting to negative energies. In spite of this, states N$'$ remain empty because of the energy barrier that separates them from the scattered states at the wire surface.  Thus the trajectory in Fig.\ \ref{f.trajectoryenergy} contributes to force only at a higher velocity $v>v_L$, when $\epsilon_3$ crosses zero and scattered quasiparticles can escape from the near region.
In Sec.\ \ref{s.hydrof} we calculate the force on a cylinder assuming the quasiparticle potential is fixed by ideal-fluid flow field (\ref{e.iflff}), $a=p_F \hat{\bm p}\cdot \bm v_s$. We indeed find that the force becomes nonzero at a velocity $v_c$ that is larger than the Landau velocity (\ref{e.landauvf}).

Another effect that has to be taken into account is that the local superfluid velocity at points P and Q (Fig.\ \ref{f.CylinderFlow}) exceeds the Landau velocity already at $v=v_L/2$. Thus depairing in the near region takes place starting from this velocity. In Fig.\ \ref{f.trajectoryenergy} this appears as scattering from states of the type L to L$'$ at velocities where $\epsilon_2$ is negative. 
This modifies the flow field around the object from the ideal flow profile (\ref{e.iflff}). 
In Secs.\ \ref{s.noc} and \ref{s.enr} we calculate the flow field at $v>v_L/2$  in two extreme cases, and estimate its effect on the drag force.
   
The theory behind all the calculations is the Fermi liquid theory of superfluidity. A review of this theory is given by Serene and Rainer \cite{Serene83}. Here we use only the low-frequency, long-wavelength limit of the general theory, which is  described in Section 7 of the review \cite{Serene83}. The quasiparticle distribution is expressed by particle-type and hole-type distribution functions $\phi_{B1}(\hat{\bm p},\bm r,\epsilon, t)$ and $\phi_{B2}(\hat{\bm p},\bm r,\epsilon, t)$, which describe excitations traveling at velocities $\bm v=\pm v_F\hat{\bm p}\sqrt{\tilde\epsilon^2-|\Delta|^2}/\tilde\epsilon$, respectively. In addition to momentum direction $\hat{\bm p}$ and energy $\epsilon$, the distributions depend on location $\bm r$ and time $t$ in the general case. We use the short-hand $\tilde\epsilon=\epsilon- a$. The excitations can also have magnetic properties, which are described by vector distribution functions $\bm\phi_{B1}$ and $\bm\phi_{B2}$ \cite{Kieselmann83}. In this study we neglect magnetic excitations. The distribution functions take values in the range $[-\frac12,\frac12]$ and their equilibrium form is $\phi_{B1}=\phi_{B2}=-\frac12\tanh(\epsilon/2T)$. 
Under a Galilei transformation to a frame moving at uniform velocity $\bm v$, the distribution functions change as $\phi_{Bi}(\hat{\bm p}, \epsilon) \to \phi_{Bi}(\hat{\bm p}, \epsilon + p_F \hat{\bm p} \cdot  \bm v)$.

We give here the equations of the low frequency dynamics in the simplified form that we use. An essential quantity is the antisymmetric quasiparticle potential $a$. In the approximation where we neglect all Fermi liquid interactions higher than first order, $F_l^s=0$ for $l\ge2$, we have $a(\hat{\bm p},\bm r, t)=\bm\alpha(\bm r, t)\cdot\hat{\bm p}$ and 
%\begin{eqnarray}
%\bm\alpha=mv_F\bm v_s+\frac12\frac{F_1^s}{1+\frac13F_1^s}\int\frac{d\Omega_p}{4\pi}\hat{\bm p}
%\int_{-E_c}^{E_c} d\epsilon\,\frac{|\tilde\epsilon|}{\sqrt{\tilde\epsilon^2-|\Delta|^2}}\theta(\tilde\epsilon^2-|\Delta|^2)(\phi_{B1}+\phi_{B2}).\label{e.atillfds2}
%\end{eqnarray}
\begin{equation}\label{e.atillfds2}
\begin{split}
\bm\alpha = m v_F \bm v_s &+ \frac{1}{2} \frac{F_1^s}{1 + \frac{1}{3} F_1^s}\int\frac{d\Omega_p}{4\pi}\hat{\bm p}
\int_{-E_c}^{E_c} d\epsilon \\
&\times \left\{\frac{|\tilde\epsilon|}{\sqrt{\tilde\epsilon^2-|\Delta|^2}}\theta(\tilde\epsilon^2-|\Delta|^2)(\phi_{B1}+\phi_{B2}) \right\}.
\end{split}
\end{equation}
Here $\int d\Omega_p$ denotes integration over the unit sphere of $\hat{\bm p}$, $E_c$ is a high energy cutoff and $\theta(x)$ is the Heaviside step function. The parameter $F_1^s$ is related to the effective mass $m^*$ by $m^*/m=1+F_1^s/3$. The distributions $\phi_{B1}$ and $\phi_{B2}$ carry independent information only on positive energies as they are related by $\phi_{B1}(\hat{\bm p},\bm r,\epsilon, t)=-\phi_{B2}(-\hat{\bm p},\bm r,-\epsilon, t)$.
As a first test of Eq.\ (\ref{e.atillfds2}), consider full equilibrium but seen from a frame moving with velocity $\bm v$.
The quasiparticle distribution is then given by $\phi_{Bi} = - \frac{1}{2} \tanh \left[ \left( \epsilon + p_F \hat{\bm p} \cdot  \bm v  \right) / 2 T  \right]$. Substituting this into Eq.\ (\ref{e.atillfds2}) gives $a = - p_F \hat{\bm p} \cdot \bm v$, which was claimed above. The second test is that at $T=0$ and $v_s<v_L$, a consistent solution of Eq.\ (\ref{e.atillfds2}) with quasiparticles in equilibrium is $a = p_F \hat{\bm p} \cdot \bm v_s$.
For known distribution functions and $a$, the
mass current density is given by
%\begin{eqnarray}
%\bm j(\bm r,t)=mv_FN(0)\int\frac{d\Omega_p}{4\pi}\hat{\bm p}
%\int_{-E_c}^{E_c} d\epsilon\,\frac{|\tilde\epsilon|}{\sqrt{\tilde\epsilon^2-|\Delta|^2}}\theta(\tilde\epsilon^2-|\Delta|^2)(\phi_{B1}+\phi_{B2}),
%\label{e.SR2.12c}\end{eqnarray}
\begin{equation}\label{e.SR2.12c}
\begin{split}
\bm j(\bm r,t) &= m v_F N(0) \int\frac{d\Omega_p}{4\pi}\hat{\bm p}
\int_{-E_c}^{E_c} d\epsilon \\
& \times \left\{ \frac{|\tilde\epsilon|}{\sqrt{\tilde\epsilon^2-|\Delta|^2}}\theta(\tilde\epsilon^2-|\Delta|^2)(\phi_{B1}+\phi_{B2}) \right\},
\end{split}
\end{equation}
where $2N(0)=m^*p_F/\pi^2\hbar^3$ is the quasiparticle density of states at the Fermi surface.  It is worth noting that also the supercurrent is contained in Eq.\ (\ref{e.SR2.12c}) through $ a$ even though the distribution functions take their equilibrium values, as discussed in connection with the semiconductor model above (Fig.\ \ref{f.BCSExcitationSpectrum}). An additional condition is that the mass current has to be conserved, which in time-independent case leads to Eq.\ (\ref{e.ndjeqo}).
Equations (\ref{e.ndjeqo}), (\ref{e.vsgp}), (\ref{e.atillfds2}) and (\ref{e.SR2.12c}) form a set that determines $\bm v_s$ at supercritical velocities for given $\phi_{B1}$, $\phi_{B2}$ and $|\Delta|$.
One more equation that determines $\Delta(\hat{\bm p},\bm r)$ is needed in general, but here we mostly assume $|\Delta|$ to be a constant, for simplicity. Once the flow field and the distribution functions are known, we can calculate the stress tensor,
%\begin{equation}\label{eq:stress_tensor}
%\tensor\Pi(\bm r, t) = v_F p_F N(0) \int\frac{d\Omega}{4\pi}\hat{\bm p}\,\hat{\bm p}\int_{-E_c}^{E_c} d\epsilon\, \,\theta(\tilde\epsilon^2-|\Delta|^2) (\phi_{B1} - \phi_{B2}).
%\end{equation}
\begin{equation}\label{eq:stress_tensor}
\begin{split}
\tensor\Pi(\bm r, t) &= v_F p_F N(0) \int\frac{d\Omega}{4\pi}\hat{\bm p}\,\hat{\bm p}\int_{-E_c}^{E_c} d\epsilon \\
&\times \left\{ \theta(\tilde\epsilon^2-|\Delta|^2) (\phi_{B1} - \phi_{B2}) \right\}.
\end{split}
\end{equation}
The force exerted on a surface with area $dA$ and normal $\hat{\bm n}$ is then given by $(\hat{\bm n} \cdot \tensor\Pi) \, dA$.

\section{Boundary condition}\label{s.bc}

In this section we introduce {\em a mesoscopic version of the diffuse boundary condition for Fermi superfluids}. The diffuse boundary condition is a commonly used model to describe the reflection of radiation or particles from a surface. The basic assumption is that the reflected radiance is independent of the direction of the incoming radiation. The combination of diffuse and specular reflection has commonly been applied to normal-state Fermi liquids. The application to the superfluid state can be more complicated, in particular if the superfluid state is modified in a surface layer. This typically occurs in non-s-wave superconductors, where the order parameter has nontrivial structure on the length scale of the superfluid coherence length $\xi_0$ from the surface. A quasiparticle reflected from the surface can be Andreev reflected back to the surface from the surface layer. The classical diffuse reflection model is insufficient to properly include the repeated Andreev and bare surface reflections, and one needs a model that works on the quantum level. See Refs.\  \cite{Kurkijarvi90,Nagai08,Shelankov00,Eschrig09} for discussion of some of these models. These quantum models give the ``dressed'' reflection of a bulk quasiparticle, where the surface layer modifies the bare reflection at the surface. The reflected distribution generally has a smooth background and peaks in the specular and retroreflection directions \cite{Kieselmann83,Zhang88}. For many problems these quantum calculations are too complicated. As an alternative, we formulate here a model that mimics the diffuse reflection on a mesoscopic scale ($>\xi_0$).
This model satisfies all the necessary conservation laws for elastic scattering. Its analytic form simplifies its application to many problems. It provides a kind of first approximation, against which more sophisticated reflection models can be compared with.
 
Our starting point is the low-frequency superfluid dynamics as described in Sec.\ \ref{s.basics}. The boundary condition can be used for arbitrary bulk gap amplitude $|\Delta (\hat{\bm p},\bm r, t)|$, either singlet or triplet. In addition to the expression for mass current (\ref{e.SR2.12c}), we need to define 
the number current density of excitations \cite{Ashauer89},
%\begin{eqnarray}
%\bm j_e(\bm r,t)=v_FN(0)\int\frac{d\Omega_p}{4\pi}\hat{\bm p}
%\int_{-E_c}^{E_c} d\epsilon\,\theta(\tilde\epsilon^2-|\Delta|^2)(\phi_{B1}-\phi_{B2}).
%\label{e.Ash2.1.25}\end{eqnarray}
\begin{equation}\label{e.Ash2.1.25}
\begin{split}
\bm j_e(\bm r,t) &= v_F N(0) \int\frac{d\Omega_p}{4\pi}\hat{\bm p}
\int_{-E_c}^{E_c} d\epsilon \\
&\times \left\{ \theta(\tilde\epsilon^2-|\Delta|^2)(\phi_{B1}-\phi_{B2}) \right\}.
\end{split}
\end{equation}

Our goal is to set up boundary condition to describe elastic reflection from a planar piece of an impenetrable wall. We study the problem in the rest frame of the wall, and denote the surface normal with $\hat{\bm n}$. The boundary condition have to obey conservation laws. Mass conservation requires that the mass current component perpendicular to the wall has to vanish, $\hat{\bm n}\cdot\bm j=0$. In addition, it is shown in Ref.\ \cite{Ashauer89} that the  excitation number current has to be conserved by elastic scattering, $\hat{\bm n}\cdot\bm j_e=0$.
In the energy representation used above, the energy conservation is automatically satisfied if the outgoing excitations are at the same energy as the incoming ones.

On the surface we define
\begin{equation}\label{e.abcdwlke}
\begin{split}
A(\epsilon) &=\int_{\hat{\bm n}\cdot\hat{\bm p}<0}d\Omega_{p}|\hat{\bm n}\cdot\hat{\bm p}|N(\hat{\bm p},\epsilon)\phi_{B1}(\hat{\bm p},\epsilon) \\
&-\int_{\hat{\bm n}\cdot\hat{\bm p}>0}d\Omega_{p}|\hat{\bm n}\cdot\hat{\bm p}|N(\hat{\bm p},\epsilon)\phi_{B2}({\hat{\bm p}},\epsilon),\\
B(\epsilon) &= \int_{\hat{\bm n}\cdot\hat{\bm p}<0}d\Omega_{p}|\hat{\bm n}\cdot\hat{\bm p}|\Theta(\hat{\bm p},\epsilon)\phi_{B1}(\hat{\bm p},\epsilon) \\
&+\int_{\hat{\bm n}\cdot\hat{\bm p}>0}d\Omega_{p}|\hat{\bm n}\cdot\hat{\bm p}|\Theta(\hat{\bm p},\epsilon)\phi_{B2}({\hat{\bm p}},\epsilon).
\end{split}
\end{equation}
Here we have dropped the parameters $\bm r$ and $t$ for simplicity, and defined
%\begin{eqnarray}
%\Theta(\hat{\bm p},\epsilon)=\theta([\epsilon- a(\hat{\bm p})]^2-|\Delta(\hat{\bm p})|^2),\quad
%N(\hat{\bm p},\epsilon)=\frac{|\epsilon- a(\hat{\bm p})|}{\sqrt{[\epsilon- a(\hat{\bm p})]^2-|\Delta(\hat{\bm p})|^2}}\theta([\epsilon- a(\hat{\bm p})]^2-|\Delta(\hat{\bm p})|^2).
%\end{eqnarray}
\begin{align}
\Theta(\hat{\bm p},\epsilon) &=\theta([\epsilon- a(\hat{\bm p})]^2-|\Delta(\hat{\bm p})|^2), \\
\nu(\hat{\bm p},\epsilon) &= \frac{|\epsilon- a(\hat{\bm p})|}{\sqrt{[\epsilon- a(\hat{\bm p})]^2-|\Delta(\hat{\bm p})|^2}}, \\
N(\hat{\bm p},\epsilon) &= \nu(\hat{\bm p},\epsilon) \Theta(\hat{\bm p},\epsilon).
\end{align}
We see that $A$ and $B$ are fully determined by the incoming excitations.
We now state the boundary condition by expressing the outgoing distributions as
\begin{equation}\label{e.mcbcsfg6}
\begin{split}
\phi_{B1}(\hat{\bm p},\hat{\bm n}\cdot\hat{\bm p}>0 ,\epsilon)
&=\frac{g(\epsilon)}{2} \left[\nu^{-1}(\hat{\bm p},\epsilon) A(\epsilon) + B(\epsilon)\right],\\
\phi_{B2}(\hat{\bm p},\hat{\bm n}\cdot\hat{\bm p}<0 ,\epsilon)
&=\frac{g(\epsilon)}{2} \left[-\nu^{-1}(\hat{\bm p},\epsilon) A(\epsilon) + B(\epsilon)\right].
\end{split}
\end{equation}
By this construction we can satisfy the conservation laws $\hat{\bm n}\cdot\bm j=\hat{\bm n}\cdot\bm j_e=0$ by fixing
\begin{eqnarray}
g^{-1}(\epsilon)=\int_{\hat{\bm n}\cdot\hat{\bm p}>0}d\Omega_{p}\hat{\bm n}\cdot\hat{\bm p}\,\Theta(\hat{\bm p},\epsilon),
\label{eq:bc_ginv}\end{eqnarray}
provided that $\hat{\bm n} \cdot \bm \alpha = 0$ and $|\Delta(\underline{\hat{\bm p}})| = |\Delta(\hat{\bm p})|$. Here $\underline{\hat{\bm p}} = \hat{\bm p} - 2 \hat{\bm n} (\hat{\bm n} \cdot \hat{\bm p})$ is the direction of specular reflection. The first condition just means that there is no flow through the surface. The second condition limits the possible forms of the gap amplitude on the surface, but it shouldn't be too restrictive. The condition is fulfilled, e.g., for s-wave superfluids and the two bulk phases of superfluid $^3$He, the A phase and the B phase.

We now study some properties of the boundary condition (\ref{e.mcbcsfg6}). It represents diffuse reflection, since the outgoing distribution depends on the incoming one through $A$ and $B$, which depend only on energy. In the normal state $|\Delta|=0$, and the boundary condition (\ref{e.mcbcsfg6}) reduces to the standard diffuse boundary condition where $g=1/\pi$. We see that $\phi_{B1}=\phi_{B2}=\phi(\epsilon)$ is a consistent solution of the boundary condition for any $a$. There is conversion between the branches, that is, an incoming particle-like excitation is reflected as hole-like excitation and vice versa. Branch conversion takes place predominantly at low energies $\epsilon\sim|\Delta|$. At higher energies the conversion becomes small because $\sqrt{\tilde\epsilon^2-|\Delta|^2}/|\tilde\epsilon|\rightarrow 1$. The conversion also vanishes for $ a=0$ and constant $|\Delta(\hat{\bm p})| = \Delta$ because in this case the factor $\sqrt{\tilde\epsilon^2-|\Delta|^2}/|\tilde\epsilon|=\sqrt{\epsilon^2-\Delta^2}/|\epsilon|$ is a function of energy only, and such factors cancel in substituting (\ref{e.abcdwlke}) into (\ref{e.mcbcsfg6}). The boundary condition (\ref{e.mcbcsfg6}) seems to be the simplest generalization of the normal-state diffusive boundary condition to superfluid state that satisfies the conditions $\hat{\bm n}\cdot\bm j=\hat{\bm n}\cdot\bm j_e=0$.

\section{A small object}\label{s.ion}

In this section we calculate the drag force on a small object using a modified version of the boundary condition introduced in Sec.\ \ref{s.bc}. We assume the object to be small compared to the coherence length, but sufficiently big that we can neglect its recoil in collisions with quasiparticles. A negative ion  in superfluid $^3$He could fall under this characterization, see Refs.\ \cite{Fetter87,Borghesani07} for reviews. Small objects have been extensively studied using the quantum approach, modeling the scattering of quasiparticles from the object by scattering phase shifts \cite{Baym79,Salomaa80b,TKR81,Shevtsov17,Tsutsumi17}. We are interested in the motion of the object at velocities exceeding the Landau critical velocity. This problem was previously studied by Bowley assuming a constant cross section \cite{Bowley77}, and by Ashauer and Rainer using the quantum approach \cite{Rainer87,Ashauer88,Ashauer89}. Ashauer and Rainer give examples of the scattered quasiparticle distributions, but do not calculate the drag force.

We study the problem in the rest frame of the object, with the object located at the origin. The diameter of the object is denoted by $d$. We assume that $d \ll \xi_0$.
The boundary condition of Sec.\ \ref{s.bc} was formulated for a piece of wall with normal $\hat{\bm n}$. In order to satisfy conservation of mass and excitation number, we required that the normal components of $\bm j$ and $\bm j_e$ vanish at the surface, $\hat{\bm n} \cdot \bm j = \hat{\bm n} \cdot \bm j_e = 0$. 
In the case of a small object we use a slightly modified approach. We are only interested in the behavior of the flow at a scale $\lambda\gg d$ where the object is essentially pointlike. We require that the conservation of mass and excitation number are satisfied at this scale, meaning that the mass flux and the excitation flux through a sphere of radius $r \sim \lambda$ around the object have to vanish, $\int r^2 d\Omega_r \hat{\bm r} \cdot \bm j = \int r^2 d\Omega_r \hat{\bm r} \cdot \bm j_e = 0$. Here $\int r^2 d\Omega_r$ denotes integration over a sphere of radius $r$, and $\hat{\bm r}$ is the radial unit vector of the spherical coordinate system.

Following closely Sec.\ \ref{s.bc}, we define coefficients $A$ and $B$ that depend on incident distributions,
\begin{equation}\label{eq:ion_AB}
\begin{split}
A(\epsilon) &=\int d\Omega_{p} N(\hat{\bm p},\epsilon) \phi_{B1}(\hat{\bm p}, -r\hat{\bm p}, \epsilon) \\
&- \int d\Omega_{p} N(\hat{\bm p},\epsilon) \phi_{B2}(\hat{\bm p}, r\hat{\bm p}, \epsilon),\\
B(\epsilon) &=\int d\Omega_{p} \Theta(\hat{\bm p},\epsilon) \phi_{B1}(\hat{\bm p}, -r\hat{\bm p}, \epsilon) \\
&+ \int d\Omega_{p} \Theta(\hat{\bm p},\epsilon) \phi_{B2}(\hat{\bm p}, r\hat{\bm p}, \epsilon).
\end{split}
\end{equation}
Note that since the object is essentially pointlike, the incident distributions with momentum direction $\hat{\bm p}$ come from directions $\hat{\bm r} = \pm \hat{\bm p}$.
We propose a boundary condition where the scattered distributions are given by
\begin{equation}\label{eq:ion_distributions}
\begin{split}
\phi_{B1}(\hat{\bm p}, r\hat{\bm p}, \epsilon) &= \frac{g(\epsilon)}{2} \left[\nu^{-1}(\hat{\bm p},\epsilon) A(\epsilon) + B(\epsilon)\right],\\
\phi_{B2}(\hat{\bm p}, -r\hat{\bm p}, \epsilon) &= \frac{g(\epsilon)}{2} \left[-\nu^{-1}(\hat{\bm p},\epsilon) A(\epsilon) + B(\epsilon)\right],
\end{split}
\end{equation}
with
\begin{equation}\label{eq:ion_geinv}
g^{-1}(\epsilon) = \int d\Omega_{p} \Theta(\hat{\bm p},\epsilon).
\end{equation}
This satisfies the conservation laws $\int r^2 d\Omega_r \hat{\bm r} \cdot \bm j = \int r^2 d\Omega_r \hat{\bm r} \cdot \bm j_e = 0$, conserves energy, and has the same properties as the boundary condition in Sec.\ \ref{s.bc}. 

To proceed, we assume that the object moves at a constant velocity $\bm v$ in the laboratory frame.
Since the object is small, it does not disturb the fluid flow nor the gap. This means that there is a uniform flow $\bm v_s = - \bm v$ in the rest frame of the object with $\bm \alpha = p_F \bm v_s$ and $|\Delta(\hat{\bm p})| = \Delta$.
Incident distributions are zero-temperature equilibrium distributions in the laboratory frame, $\phi_{B1}(\hat{\bm p}, -r\hat{\bm p}, \epsilon) = \phi_{B2}(\hat{\bm p}, r\hat{\bm p}, \epsilon) = 1/2 - \theta(\epsilon + p_F \bm v \cdot \hat{\bm p})$.
%These are obtained by Galilei transforming the zero-temperature equilibrium distribution function $1/2 - \theta(\epsilon)$ from the laboratory frame to the object frame.

Calculating the integrals in Eq.\ (\ref{eq:ion_AB}) we see that $A(\epsilon) = 0$ and
\begin{equation}
\frac{g(\epsilon) B(\epsilon) }{2} = 
\begin{cases}
\frac{1}{2} \frac{\epsilon}{\Delta - \alpha} & \text{when } |\epsilon| \leq \alpha - \Delta \\
\frac{1}{2} - \theta(\epsilon) & \text{when } |\epsilon| > \alpha - \Delta
\end{cases}.
\end{equation}
To calculate the force $\bm F$ exerted on the object, we integrate the radial component of the stress tensor (\ref{eq:stress_tensor}) over a sphere of radius $r \sim \lambda$ centered at the origin, resulting in
\begin{equation}\label{eq:ion_force}
\bm F = \theta(v - v_L) \frac{\left( v - v_L  \right)^2 \left( v + v_L  \right)}{v^3} \bm F_n.
\end{equation}
Here $\bm F_n = - p_F n_f \sigma \bm v$ is the force in the normal state \cite{Virtanen06}, $n_f$ is the number density of fermions, and $\sigma$ is the cross-section of the particle. Similar calculation, but apparently with somewhat different assumptions, was made by Bowley \cite{Bowley77}. His result is smaller than the one in Eq.\ (\ref{eq:ion_force}) by a factor of $1-v_L/v$.

\begin{figure}[tb] %  figure placement: top, bottom, (not here or page)
\centering
\includegraphics[width=0.8\linewidth]{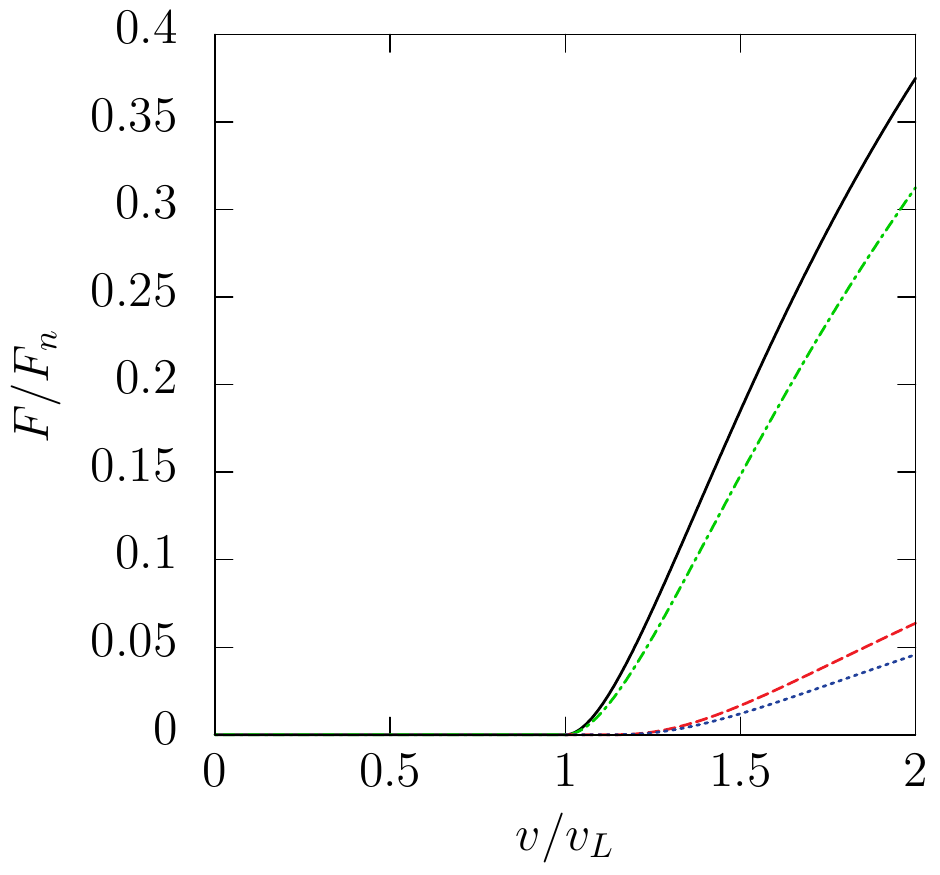} 
\caption{Drag force as a function of velocity for a small pointlike object (diameter $\ll \xi_0$, see Sec.\ \ref{s.ion}) and for a macroscopic cylinder (diameter $\gg \xi_0$, see Sec.\ \ref{s.hydrof}). The solid black line (dashed red line) represents the force exerted on the pointlike object (macroscopic cylinder) assuming that the scattered distributions are given by the diffuse boundary condition, Eq.\ (\ref{eq:ion_distributions}) [Eq.\ (\ref{e.mcbcsfg6})]. The dot-dashed green line (dotted blue line) represents the force exerted on the pointlike object (macroscopic cylinder) assuming that the scattered excitations are given by equilibrium distributions in the object frame, $1 / 2 - \theta(\epsilon)$. We have assumed zero temperature, constant gap, no collisions between quasiparticles, and ideal flow field. For both objects there is a clear critical velocity. For the pointlike object the critical velocity equals the Landau velocity. In the case of the cylinder the critical velocity is shifted to $\sim 1.12 v_L$ due to the spatial dependence of the flow field. At high velocities $v \gg v_L$ the forces approach the normal state values.}
\label{fig:IonHydroForce}
\end{figure}

The force in Eq.\ (\ref{eq:ion_force}) vanishes at velocities lower than $v_L$ since the object cannot scatter quasiparticles. At velocities much higher than the critical velocity $v_L$ the force approaches the normal state value. Figure \ref{fig:IonHydroForce} shows the ratio $F / F_n$ as a function of velocity. 
As a comparison, the figure also shows the force calculated in a case where we don't use the boundary condition (\ref{eq:ion_distributions}), but instead assume that the scattered distributions are equilibrium distributions in the object frame, $\phi_{B1}(\hat{\bm p}, r\hat{\bm p}, \epsilon) = \phi_{B2}(\hat{\bm p}, -r\hat{\bm p}, \epsilon) = 1 / 2 - \theta(\epsilon)$. 
We see that the qualitative behavior is similar in both cases. Critical velocities are equal, and the results agree near the critical velocity. At larger velocities the diffuse boundary condition (\ref{eq:ion_distributions}) yields a slightly larger force.

\section{A large object in the collisionless approximation}\label{s.hydrof}

In this section we apply the boundary condition of Sec.\ \ref{s.bc} to a cylinder of finite radius $R \gg \xi_0$ moving at constant velocity $\bm v = v \hat{\bm x}$ perpendicular to its axis $\hat{\bm z}$. We assume zero temperature $T = 0$ and no collisions between quasiparticles. We use the boundary condition to calculate the stress tensor for a given flow field. Finally, we use the results to calculate the drag force exerted on the cylinder assuming an ideal flow field around the cylinder. The purpose of this calculation is to demonstrate that even though the cylinder scatters quasiparticles at velocities $v > v_L/2$, the spatially varying flow field can prevent them from escaping from the vicinity of the cylinder, thus reducing the force significantly.

We shall work in the rest frame of the cylinder. Because there are no collisions between excitations, the distribution functions satisfy \cite{Serene83} $\hat{\bm p} \cdot \bm \nabla \phi_{Bi}(\hat{\bm p}, \bm r, \epsilon) = 0$. Inside the gap, i.e., at points where $[\epsilon - a(\hat{\bm p}, \bm r)]^2 - |\Delta(\hat{\bm p}, \bm r)|^2 < 0$, the distribution  functions are not defined. Thus the distribution functions are piecewise constant along trajectories $\bm r = \bm r_0 + s \hat{\bm p}$, where $\bm r_0$ is fixed and $s \in \mathbb{R}$. Andreev reflection occurs at points $[\epsilon - a(\hat{\bm p}, \bm r)]^2 - |\Delta(\hat{\bm p}, \bm r)|^2 = 0$. At these points the distribution functions are equal, $\phi_{B1}(\hat{\bm p}, \bm r, \epsilon) = \phi_{B2}(\hat{\bm p}, \bm r, \epsilon)$.

In order to determine the scattered distributions (\ref{e.mcbcsfg6}), we need to calculate the coefficients $A$, $B$, and $g$ from Eqs.\ (\ref{e.abcdwlke}) and (\ref{eq:bc_ginv}).
We split the integrals over $\hat{\bm p}$ into two parts, over free states and over bound states. Free states are such that, at given $\epsilon$, no Andreev reflection occurs at a trajectory along $\hat{\bm p}$. In Fig.\ \ref{f.trajectoryenergy} these correspond to regions M and M$'$. This means that the value of the incident distribution at the surface is equal to the value of the distribution at the far region, which we take to be the equilibrium distribution in the laboratory frame,
\begin{equation}
\begin{split}
\phi_{B1}^{\text{free}}(\hat{\bm p}, \hat{\bm n}\cdot\hat{\bm p} < 0, \epsilon)& = \phi_{B2}^{\text{free}}(\hat{\bm p}, \hat{\bm n}\cdot\hat{\bm p} > 0, \epsilon) \\
&= \phi_{Bi}^{\infty}(\hat{\bm p},\epsilon) = \frac{1}{2} - \theta(\epsilon + p_F \bm v \cdot \hat{\bm p} ).
\end{split}
\end{equation}
Bound states, on the other hand, are such that Andreev reflection occurs at some points of a trajectory along $\hat{\bm p}$. In Fig.\ \ref{f.trajectoryenergy} these correspond to regions L and L$'$. Since the distributions $\phi_{B1}$ and $\phi_{B2}$ are equal at the point of Andreev reflection, the incident distribution $\phi_{B1}$ is equal to the scattered distribution $\phi_{B2}$, and vice versa,
\begin{equation}
\begin{split}
\phi_{B1}^{\text{bound}}(\hat{\bm p}, \hat{\bm n}\cdot\hat{\bm p} < 0, \epsilon) &= \frac{g(\epsilon)}{2} \left[-\nu^{-1}(\hat{\bm p},\epsilon) A(\epsilon) + B(\epsilon)\right], \\
\phi_{B2}^{\text{bound}}(\hat{\bm p}, \hat{\bm n}\cdot\hat{\bm p} > 0, \epsilon) &= \frac{g(\epsilon)}{2} \left[\nu^{-1}(\hat{\bm p},\epsilon) A(\epsilon) + B(\epsilon)\right].
\end{split}
\end{equation}
Note that the regions K, K$'$, N and N$'$ in Fig.\ \ref{f.trajectoryenergy} are not relevant here, since we are discussing points on the surface of the cylinder.

Inserting the incident distributions into Eq.\ (\ref{e.abcdwlke}) leads to a self-consistency equation for $A$ and $B$,

\begin{widetext}

\begin{equation}\label{eq:AB_self-consistency}
\begin{split}
\left[ - 4 g^{-1} + \int_{\text{free}} d\Omega_{p} |\hat{\bm n} \cdot \hat{\bm p}| \Theta(\hat{\bm p},\epsilon)  \right] \frac{g A}{2} + \left[ \int_{\text{free}} d\Omega_{p} (\hat{\bm n} \cdot \hat{\bm p}) N(\hat{\bm p},\epsilon)  \right] \frac{g B}{2} &= \int_{\text{free}} d\Omega_{p} (\hat{\bm n} \cdot \hat{\bm p}) N(\hat{\bm p},\epsilon) \phi_{Bi}^{\infty}(\hat{\bm p},\epsilon)  , \\
\left[ \int_{\text{free}} d\Omega_{p} (\hat{\bm n} \cdot \hat{\bm p}) \nu^{-1}(\hat{\bm p},\epsilon) \Theta(\hat{\bm p},\epsilon) \right] \frac{g A}{2} + \left[ \int_{ \text{free}} d\Omega_{p} |\hat{\bm n} \cdot \hat{\bm p}| \Theta(\hat{\bm p},\epsilon) \right] \frac{g B}{2} &= \int_{\text{free}} d\Omega_{p} |\hat{\bm n} \cdot \hat{\bm p}| \Theta(\hat{\bm p},\epsilon) \phi_{Bi}^{\infty}(\hat{\bm p},\epsilon)  .
\end{split}
\end{equation}
\end{widetext}
Here we have transformed the integrals over bound states into integrals over free states using the identity $\int d\Omega_{p} = \int_{\text{free}} d\Omega_{p} + \int_{\text{bound}} d\Omega_{p}$.
This is a linear system of two equations and two unknowns, but the coefficients are quite complicated. What can we say about the solution?

Let us consider a fixed point $\bm r_0$ on the surface of the cylinder. 
Free states satisfy $[\epsilon - a(\hat{\bm p}, \bm r)]^2 - |\Delta(\hat{\bm p}, \bm r)|^2 > 0$ along the whole trajectory $\bm r = \bm r_0 + s \hat{\bm p}$. Here $s \geq 0$ if $\hat{\bm n} \cdot \hat{\bm p} > 0$ and $s \leq 0$ if $\hat{\bm n} \cdot \hat{\bm p} < 0$. 
Let us denote $E^{\pm}(\hat{\bm p}, \bm r) = a(\hat{\bm p}, \bm r) \pm |\Delta(\hat{\bm p}, \bm r)|$. In addition, let us denote the maximum and the minimum of $E^{\pm}(\hat{\bm p}, \bm r)$ along the trajectory by $E^{\pm}_{\text{max}}(\hat{\bm p}, \bm r_0)$ and $E^{\pm}_{\text{min}}(\hat{\bm p}, \bm r_0)$, respectively. We can then split the free states into two categories, $F_1 = \left\{ \hat{\bm p} \, \vert \, E^{+}_{\text{max}}(\hat{\bm p}, \bm r_0) < \epsilon \right\}$ and $F_2 = \left\{ \hat{\bm p} \, \vert \, E^{-}_{\text{min}}(\hat{\bm p}, \bm r_0) > \epsilon  \right\}$.

If both $F_1$ and $F_2$ are empty, Eq.\ (\ref{eq:AB_self-consistency}) tells us that $A = 0$, but leaves $B$ unspecified. Since all states are bound, it is natural to assume that they are in equilibrium with the cylinder. We therefore have $g A / 2 = 0$ and $g B / 2 = 1 / 2 - \theta(\epsilon)$.
If either $F_1$ or $F_2$, but not both, is empty, we see that the solution to Eq.\ (\ref{eq:AB_self-consistency}) is $g A / 2 = 0$ and $g B / 2 = 1 / 2 - \theta(\epsilon)$. 
If neither $F_1$ nor $F_2$ is empty, then we need to calculate the coefficients in Eq.\ (\ref{eq:AB_self-consistency}) numerically.

Let us study the regions $F_1$ and $F_2$ more closely.
Region $F_1$ is empty when $\min_{\hat{\bm p}} \left\{ E^{+}_{\text{max}}(\hat{\bm p}, \bm r_0) \right\} \geq \epsilon $. 
Region $F_2$ is empty when $\max_{\hat{\bm p}} \left\{ E^{-}_{\text{min}}(\hat{\bm p}, \bm r_0) \right\} \leq  \epsilon $. For pure singlet or pure triplet superfluid we have $E^{-}(-\hat{\bm p}, \bm r) = - E^{+}(\hat{\bm p}, \bm r)$. This means that region $F_2$ is empty when $\min_{\hat{\bm p}} \left\{ E^{+}_{\text{max}}(\hat{\bm p}, \bm r_0) \right\} \geq - \epsilon $.
Thus neither of the regions is empty when $|\epsilon| < -\min_{\hat{\bm p}} \left\{ E^{+}_{\text{max}}(\hat{\bm p}, \bm r_0) \right\} $.
If we define
%\begin{equation}\label{eq:bc_energy}
%\mathcal{E}(\bm r_0) = 
%\begin{cases}
%0 &  , \min_{\hat{\bm p}} \left\{ E^{+}_{\text{max}}(\hat{\bm p}, \bm r_0) \right\} \geq 0 \\
%- \min_{\hat{\bm p}} \left\{ E^{+}_{\text{max}}(\hat{\bm p}, \bm r_0) \right\} & , \min_{\hat{\bm p}} \left\{ E^{+}_{\text{max}}(\hat{\bm p}, \bm r_0) \right\} < 0
%\end{cases},
%\end{equation}
\begin{equation}\label{eq:bc_energy}
\mathcal{E}(\bm r_0) = - \min_{\hat{\bm p}} \left\{ E^{+}_{\text{max}}(\hat{\bm p}, \bm r_0) \right\} \theta \left(  - \min_{\hat{\bm p}} \left\{ E^{+}_{\text{max}}(\hat{\bm p}, \bm r_0) \right\}  \right),
\end{equation}
then the only non-trivial region of energies where we need to calculate $A$ and $B$ numerically is $|\epsilon| < \mathcal{E}$. When $|\epsilon| \geq \mathcal{E}$, the scattered distributions are $\phi_{B1}(\hat{\bm p},\hat{\bm n}\cdot\hat{\bm p}>0 ,\epsilon) = \phi_{B2}(\hat{\bm p},\hat{\bm n}\cdot\hat{\bm p}<0 ,\epsilon) = 1/2 - \theta(\epsilon)$.
%Note that $\mathcal{E}$ vanishes when $v < v_L$. It can also vanish for velocities larger than $v_L$, depending on the spatial dependence of the $\bm \alpha$-field.
Physically the fact that $\mathcal{E}$ is zero means that no quasiparticles can escape from the vicinity of the cylinder. 
%Using $\mathcal{E}$ we can write the scattered distributions as
%\begin{equation}
%\begin{split}
%\phi_{B1}(\hat{\bm p},\hat{\bm n}\cdot\hat{\bm p}>0 ,\epsilon) &=
%\begin{cases}
%\nu^{-1}(\hat{\bm p}, \epsilon) g(\epsilon)A(\epsilon) / 2 + g(\epsilon)B(\epsilon) / 2 & \text{when }  |\epsilon| < \mathcal{E} \\
%1 / 2 - \theta(\epsilon) & \text{when  }  |\epsilon| \geq \mathcal{E}
%\end{cases} \\
%\phi_{B2}(\hat{\bm p},\hat{\bm n}\cdot\hat{\bm p}<0 ,\epsilon) &=
%\begin{cases}
%- \nu^{-1}(\hat{\bm p}, \epsilon) g(\epsilon)A(\epsilon) / 2 + g(\epsilon)B(\epsilon) / 2 & \text{when }  |\epsilon| < \mathcal{E} \\
%1 / 2 - \theta(\epsilon) & \text{when }  |\epsilon| \geq \mathcal{E}
%\end{cases}.
%\end{split}
%\end{equation}

The drag force exerted on the cylinder is given by
\begin{equation}\label{eq:collisionless_force}
\bm F = l \int_{-\pi}^{\pi} R d\varphi \hat{\bm n} \cdot \tensor \Pi(R, \varphi).
\end{equation}
Here $l$ is the length of the cylinder and $\varphi$ is the azimuthal angle around the cylinder. 
The stress tensor $\tensor \Pi$ is given by Eq.\ (\ref{eq:stress_tensor}). We split the integral over $\hat{\bm p}$ again into two parts, over free states and over bound states. Substituting the incident and scattered distributions into the integrand yields
%\begin{equation}\label{eq:collisionless_stress_tensor}
%\begin{split}
%\tensor \Pi(R, \varphi) &= 2 v_F p_F N(0) \\
%\int_{0}^{\mathcal{E}} d\epsilon \bigg\{ \frac{g A}{2} &\int_{\text{free}} \frac{d \Omega_p}{4 \pi} \hat{\bm p} \hat{\bm p} \, \nu^{-1}(\hat{\bm p},\epsilon) \Theta(\hat{\bm p},\epsilon) \\
%+  \frac{g B}{2} &\int_{\text{free}} \frac{d \Omega_p}{4 \pi} \hat{\bm p} \hat{\bm p} \, \text{sgn} ( \hat{\bm n} \cdot \hat{\bm p} ) \Theta(\hat{\bm p}, \epsilon) \\
%-&\int_{\text{free}} \frac{d \Omega_p}{4 \pi} \hat{\bm p} \hat{\bm p} \, \text{sgn} ( \hat{\bm n} \cdot \hat{\bm p} ) \Theta(\hat{\bm p}, \epsilon) \phi_{Bi}^{\infty}(\hat{\bm p},\epsilon)  \bigg\}.
%\end{split}
%\end{equation}
\begin{equation}\label{eq:collisionless_stress_tensor}
\begin{split}
\tensor \Pi(R, \varphi) &= 2 v_F p_F N(0) \int_{0}^{\mathcal{E}} d\epsilon \int_{\text{free}} \frac{d \Omega_p}{4 \pi} \hat{\bm p} \hat{\bm p} \Theta(\hat{\bm p},\epsilon) \\ 
&\times \bigg\{ \frac{g(\epsilon) A(\epsilon) }{2}  \nu^{-1}(\hat{\bm p},\epsilon)  +  \frac{g(\epsilon)  B(\epsilon) }{2}  \text{sgn} ( \hat{\bm n} \cdot \hat{\bm p} )  \\
&\hspace{97pt}- \text{sgn} ( \hat{\bm n} \cdot \hat{\bm p} ) \phi_{Bi}^{\infty}(\hat{\bm p},\epsilon)  \bigg\}.
\end{split}
\end{equation}
Due to symmetries, the force is purely opposite to the direction of motion, i.e.\ $\bm F = - F \hat{\bm v}$. 
We see that the force vanishes when $\mathcal{E} = 0$ at all points on the surface of the cylinder. This means that the critical velocity can be defined as the smallest velocity for which $\mathcal{E} > 0$.

We conclude this section by applying the above results to the case of ideal flow around the cylinder. The ideal velocity field is given by Eqs.\ (\ref{e.vsgp}) and (\ref{e.iflff}), while the $\bm \alpha$-field is given by $\bm \alpha = p_F \bm v_s$. We also assume that the gap amplitude is constant, $|\Delta(\hat{\bm p}, \bm r)| = \Delta$. Note that the ideal flow field is not consistent with Eqs.\ (\ref{e.ndjeqo}), (\ref{e.vsgp}), (\ref{e.atillfds2}) and (\ref{e.SR2.12c}) at velocities $v > v_L / 2$. It is, however, a reasonable starting point. Self-consistent flow will be considered in Secs.\ \ref{s.noc} and \ref{s.enr}. 

Figure \ref{fig:IonHydroForce} shows the force as a function of $v$ in the case of the ideal flow. The unit of force is chosen to be the normal-state value \cite{Virtanen06} $F_n = \frac{43 \pi}{48} p_F n_f v l R$. We see that the critical velocity is increased from $v_L$, and is now approximately $1.12 v_L$. 
This is slightly smaller than $8 v_L / 7$, which is when $\epsilon_3$ crosses zero (at $s = \sqrt{3} R$) on a trajectory along $\hat{\bm p} = \hat{\bm v}$ starting from either P or Q (see Figs.\ \ref{f.CylinderFlow} and \ref{f.trajectoryenergy}). For velocities $v < 2 v_L$ the ratio $F / F_n$ is nearly an order of magnitude smaller than in the case of the small object. This shows the importance of the spatial variation of the flow field. At high velocities the force approaches the normal state value. 
It is again interesting to compare the force we obtained above with the force that is obtained if, instead of the boundary condition (\ref{e.mcbcsfg6}), we assume that the scattered distributions are equilibrium distributions in the object frame, $\phi_{B1}(\hat{\bm p}, \hat{\bm n}\cdot\hat{\bm p} > 0, \epsilon) = \phi_{B2}(\hat{\bm p}, \hat{\bm n}\cdot\hat{\bm p} < 0, \epsilon) = 1/2 - \theta(\epsilon)$. This is also shown in Fig.\ \ref{fig:IonHydroForce}. 
As in the case of the small object, the two different boundary conditions lead to a qualitatively similar force. 
Both boundary conditions yield the same critical velocity. The results agree near the critical velocity, but start to deviate slightly from each other at larger velocities where the diffuse boundary condition (\ref{e.mcbcsfg6}) produces larger force.

\section{Self-consistent flow in the collisionless approximation}\label{s.noc}

We shall now study how the excitations modify the ideal-fluid flow field around the cylinder. We assume that there are no collisions between quasiparticles. We also assume zero temperature $T = 0$ and constant gap $|\Delta(\hat{\bm p}, \bm r)| = \Delta$.

Let us define
\begin{equation}\label{eq:collisionless_integral}
\begin{split}
\bm I(\bm r) &= \int \frac{d\Omega_p}{4 \pi} \hat{\bm p} \int_{a(\hat{\bm p}, \bm r) + \Delta}^{E_c} d\epsilon N(\hat{\bm p}, \epsilon, \bm r) \\
& \times \left\{ \left( \phi_{B1}(\hat{\bm p}, \epsilon, \bm r) + \frac{1}{2} \right) + \left( \phi_{B2}(\hat{\bm p}, \epsilon, \bm r) + \frac{1}{2} \right)  \right\}.
\end{split}
\end{equation}
Using Eqs.\ (\ref{e.atillfds2}) and (\ref{e.SR2.12c}) we can write the mass current density $\bm j$ as a sum of two parts,
\begin{equation}\label{eq:collisionless_current}
\bm j = m n_f \bm v_s + \frac{3 m^* n_f}{p_F} \bm I.
\end{equation}
The first part here is explicitly proportional to the superfluid velocity, while the second part depends on excitations. Indeed, if there are no excitations present, then $\phi_{Bi} = - 1 /2 $, and thus $\bm I = 0$.
The set of equations (\ref{e.ndjeqo}), (\ref{e.vsgp}), (\ref{e.atillfds2}) and (\ref{e.SR2.12c}) that determine the flow can be written as
\begin{align}
\bm \alpha &= p_F \bm v_s + F_1^s \bm I, \label{eq:collisionless_alpha_eq} \\
\nabla^2 \psi &= - \frac{2 m}{\hbar p_F}  (3 + F_1^s) \bm \nabla \cdot \bm I, \label{eq:collisionless_psi_eq} \\
\bm v_s &= \frac{\hbar}{2 m} \bm \nabla \psi. \label{eq:collisionless_vs}
\end{align}

The integral in (\ref{eq:collisionless_integral}) is calculated over the upper branches of states in Figs.\ \ref{f.BCSExcitationSpectrum}, \ref{f.BCSExcitationSpectrumC} and \ref{f.trajectoryenergy}, denoted by blue color. In order to carry out the integration, we need to know the distribution functions $\phi_{Bi}$ for these states.
Let us consider a fixed point $\bm r_0$ in the fluid. Since we assumed that there are no collisions between quasiparticles, the distribution functions $\phi_{Bi}(\hat{\bm p}, \epsilon, \bm r)$ are piecewise constant along trajectories $\bm r = \bm r_0 + s \hat{\bm p}$, $s \in \mathbb{R}$, as we saw earlier. 

If the excitations originate from the far region, then we assume equilibrium in the laboratory frame, $\phi_{Bi} = \phi_{Bi}^{\infty} = 1 / 2 - \theta(\epsilon + p_F \bm v \cdot \hat{\bm p} )$. 

If the excitations originate from the surface of the cylinder, then the distributions are determined by the boundary condition (\ref{e.mcbcsfg6}), $\phi_{Bi} = \phi_{Bi}^{\text{bc}}$. We saw in Sec.\ \ref{s.hydrof} that when the velocity of the cylinder is below the critical velocity, the scattered distributions are $\phi_{Bi}^{\text{bc}} = 1/2 - \theta(\epsilon)$. At higher velocities the scattered distributions are smoothed, and their widths are given by $2 \mathcal{E}$. 
The corrections to $\bm I$ caused by smoothing of the scattered distributions are of order $\mathcal{E}$. This means that at velocities near the critical velocity, where $\mathcal{E}$ is small, the approximation $\phi_{Bi}^{\text{bc}} \approx 1/2 - \theta(\epsilon)$ is a decent one. We already saw this earlier when we calculated the force exerted on the small object and on the cylinder assuming ideal flow, see Fig.\ \ref{fig:IonHydroForce}. We shall therefore approximate $\phi_{Bi}^{\text{bc}} = 1/2 - \theta(\epsilon)$ in order to simplify the calculations.

Finally, it is possible that there are excitations trapped in the fluid, $\phi_{Bi} = \phi_{Bi}^{\text{trap}}$. These excitations cannot reach either the far region or the surface of the cylinder, but are instead localized somewhere in the flow field, bouncing back and forth due to repeated Andreev reflections.
In this case we consider two different models.

In \emph{Model 1} we assume that the excitations are in equilibrium with the local flow, $\phi_{Bi}^{\text{trap}} = 1 / 2 - \theta( \epsilon - a)$, meaning that the states are always empty, since $\epsilon \geq a + \Delta$. The reasoning behind this model is that, once the equilibrium is reached, there is no way for the excitations to scatter into these states. 

In \emph{Model 2} we assume two different distributions depending on $\hat{\bm p}$. On trajectories that do not intersect the cylinder the excitations are still in equilibrium with the local flow, $\phi_{Bi}^{\text{trap}} = 1 / 2 - \theta( \epsilon - a)$. On trajectories that do intersect the cylinder however, we assume that the excitations are in equilibrium with the cylinder,  $\phi_{Bi}^{\text{trap}} = 1 / 2 - \theta( \epsilon )$. The reasoning behind this model is the following. In the experiment the wire starts from rest and is accelerated until it reaches velocity $v$. At velocities below $v_L / 2$ there are no excitations present anywhere, since the local flow velocity is below $v_L$ everywhere. At higher velocities excitations start to emerge from regions on the surface of the wire where the flow velocity exceeds $v_L$. These excitations, in turn, modify the flow near the wire. It could then be possible that the flow field is modified in such a way that some of these excitations get trapped in the fluid. Note that since the excitations originate from the surface of the cylinder, their $\hat{\bm p}$ points either towards or away from the cylinder, and their distributions are determined by the boundary condition, which we assumed to be the equilibrium distribution $1/2 - \theta(\epsilon)$. 

In reality the acceleration of the wire is a complicated process which should be modelled dynamically, but here we consider these two simple extremes.
A third model, where we assume that all of the excitations trapped in the fluid are in equilibrium with the cylinder, $\phi_{Bi}^{\text{trap}} = 1 / 2 - \theta( \epsilon )$, is considered in Sec.\ \ref{s.enr}. This equilibrium could be achieved through quasiparticle-quasiparticle collisions.

At fixed $\bm r$ and $\hat{\bm p}$ we split the integration region over energy into sets M$'$, L$'$, N$'$, K$'$ according to Fig.\ \ref{f.trajectoryenergy}. In regions M$'$, L$'$ and N$'$ the excitations originate either from the far region or the surface of the cylinder, depending on $\hat{\bm p}$ and the type of the excitation ($B1$ or $B2$). In region K$'$ the excitations are trapped in the fluid. Substituting the distributions into Eq.\ (\ref{eq:collisionless_integral}) yields
\begin{widetext}
\begin{equation}\label{eq:collisionless_integral_simplified}
\begin{split}
\bm I(\bm r) &= \int_{\text{hit}+} \frac{d\Omega_p}{4 \pi} \hat{\bm p} \left\{  \int_{\epsilon_{\text{max}}}^{E_c} d\epsilon N(\hat{\bm p}, \epsilon, \bm r)  \left( \phi_{B1}^{\text{bc}}(\hat{\bm p}, \epsilon, \bm r) + \frac{1}{2} \right) + 2 \int_{\epsilon_2}^{\epsilon_{\text{max}}} d\epsilon N(\hat{\bm p}, \epsilon, \bm r)  \left( \phi_{B1}^{\text{bc}}(\hat{\bm p}, \epsilon, \bm r) + \frac{1}{2} \right) \right\} \\
&+ \int_{\text{hit}-} \frac{d\Omega_p}{4 \pi} \hat{\bm p} \left\{  \int_{\epsilon_{\text{max}}}^{E_c} d\epsilon N(\hat{\bm p}, \epsilon, \bm r)  \left( \phi_{B2}^{\text{bc}}(\hat{\bm p}, \epsilon, \bm r) + \frac{1}{2} \right) + 2 \int_{\epsilon_2}^{\epsilon_{\text{max}}} d\epsilon N(\hat{\bm p}, \epsilon, \bm r)  \left( \phi_{B2}^{\text{bc}}(\hat{\bm p}, \epsilon, \bm r) + \frac{1}{2} \right) \right\} \\
&+ \int_{\text{hit}} \frac{d\Omega_p}{4 \pi} \hat{\bm p} \left\{  2\int_{\epsilon_0}^{\epsilon_{\text{min}}} d\epsilon N(\hat{\bm p}, \epsilon, \bm r)  \left( \phi_{Bi}^{\text{trap}}(\hat{\bm p}, \epsilon, \bm r) + \frac{1}{2} \right) \right\}
\end{split}
\end{equation}
\end{widetext}
Here $\int_{\text{hit}+}$ ($\int_{\text{hit}-}$) means integration over trajectories that intersect the cylinder with $\hat{\bm n} \cdot \hat{\bm p} > 0$ ($\hat{\bm n} \cdot \hat{\bm p} < 0$), and $\int_{\text{hit}} = \int_{\text{hit}+} +\int_{\text{hit}-}$. 
The limits of energy integration are defined as $\epsilon_0(\hat{\bm p}, \bm r) = a(\hat{\bm p}, \bm r) + \Delta$, $\epsilon_2(\hat{\bm p}, \bm r) =  a_{\text{max} < }(\hat{\bm p}, \bm r) + \Delta$, $\epsilon_3(\hat{\bm p}, \bm r) =  a_{\text{max} > }(\hat{\bm p}, \bm r) + \Delta$, $\epsilon_{\text{min}} = \min \{ \epsilon_2, \epsilon_3  \}$, and $\epsilon_{\text{max}} = \max \{ \epsilon_2, \epsilon_3  \}$.
Here $ a_{\text{max} < }(\hat{\bm p}, \bm r_0)$ [$ a_{\text{max} > }(\hat{\bm p}, \bm r_0)$] is the maximum of $a(\hat{\bm p}, \bm r)$ towards (away from) the cylinder along $\bm r = \bm r_0 + s \hat{\bm p}$.

Let us now consider each of the models separately. In Model 1 we denote $\bm I = \bm I_1$ and substitute $\phi_{Bi}^{\text{bc}} = 1/2  - \theta(\epsilon)$, $\phi_{Bi}^{\text{trap}} = 1/2 - \theta(\epsilon - a)$ into Eq.\ (\ref{eq:collisionless_integral_simplified}). In Model 2 we denote $\bm I = \bm I_2$ and substitute $\phi_{Bi}^{\text{bc}} = 1/2  - \theta(\epsilon)$, $\phi_{Bi}^{\text{trap}} = 1/2 - \theta(\epsilon )$ into Eq.\ (\ref{eq:collisionless_integral_simplified}). This yields
\begin{widetext}
\begin{align}
\bm I_1(\bm r) = &\int_{\text{hit}} \frac{d\Omega_p}{2 \pi} \hat{\bm p} \, \theta \left[ - a_{\text{max} < }(\hat{\bm p}, \bm r) - \Delta  \right] \left\{ \sqrt{a(\hat{\bm p}, \bm r)^2 - \Delta^2} - \sqrt{\left[ a_{\text{max} < }(\hat{\bm p}, \bm r) - a(\hat{\bm p}, \bm r) + \Delta  \right]^2 - \Delta^2 }  \right\} \nonumber \\
-&\int_{\text{hit}} \frac{d\Omega_p}{4 \pi} \hat{\bm p} \, \theta \left[ - a_{\text{max} }(\hat{\bm p}, \bm r) - \Delta  \right] \left\{ \sqrt{a(\hat{\bm p}, \bm r)^2 - \Delta^2} - \sqrt{\left[ a_{\text{max} }(\hat{\bm p}, \bm r) - a(\hat{\bm p}, \bm r) + \Delta  \right]^2 - \Delta^2 }  \right\}, \label{eq:collisionless_integral_model_1} \\
\bm I_2(\bm r) = &\int_{\text{hit}} \frac{d\Omega_p}{2 \pi} \hat{\bm p} \, \theta \left[ - a(\hat{\bm p}, \bm r) - \Delta  \right] \left\{ \sqrt{a(\hat{\bm p}, \bm r)^2 - \Delta^2} \right\} \nonumber \\
-&\int_{\text{hit}} \frac{d\Omega_p}{2 \pi} \hat{\bm p} \, \theta \left[ - a_{\text{max} > }(\hat{\bm p}, \bm r) - \Delta  \right] \left\{ \sqrt{a(\hat{\bm p}, \bm r)^2 - \Delta^2} - \sqrt{\left[ a_{\text{max} > }(\hat{\bm p}, \bm r) - a(\hat{\bm p}, \bm r) + \Delta  \right]^2 - \Delta^2 }  \right\} \nonumber \\
+&\int_{\text{hit}} \frac{d\Omega_p}{4 \pi} \hat{\bm p} \, \theta \left[ - a_{\text{max} }(\hat{\bm p}, \bm r) - \Delta  \right] \left\{ \sqrt{a(\hat{\bm p}, \bm r)^2 - \Delta^2} - \sqrt{\left[ a_{\text{max} }(\hat{\bm p}, \bm r) - a(\hat{\bm p}, \bm r) + \Delta  \right]^2 - \Delta^2 }  \right\}. \label{eq:collisionless_integral_model_2}
\end{align}
\end{widetext}

In numerical calculations we measure $\bm \alpha$, $\bm I$, $\Delta$, and $\epsilon$ in units of $p_F v$, $\bm v_s$ in units of $v$, $\bm r$ in units of $R$, and $\psi$ in units of $2 m v R/\hbar$. In these units the flow equations (\ref{eq:collisionless_alpha_eq}), (\ref{eq:collisionless_psi_eq}) and (\ref{eq:collisionless_vs}) may be written as
\begin{align}
\bm\alpha &= \delta\bm v_s + \bm v_0+ F_1^s \bm I, \label{eq:num_alpha} \\
\nabla^2 \delta\psi &= -(3 + F_1^s)\bm\nabla\cdot \bm I, \label{eq:num_poissons} \\ 
\bm\nabla\delta\psi &= \delta\bm v_s,
\end{align}
where the field $\delta\bm v_s = \bm v_s - \bm v_0$ represents the modification to the ideal flow $\bm v_0$ [see Eqs.\ (\ref{e.vsgp}) and (\ref{e.iflff})] caused by the presence of excitations. For a given $\delta\bm v_s$ the field $\bm\alpha$ can be solved from Eq.\ (\ref{eq:num_alpha}). Calculation of the integral $\bm I$ at location $\bm r_0$ requires a search for directions and energy ranges in which excitations are able to escape the cylinder surface into the surrounding liquid. These depend on the value of $\bm\alpha$ along trajectories $\bm r = \bm r_0 + s\hat {\bm p}$ that intersect the cylinder surface. In the discrete version of Eq.\ (\ref{eq:num_alpha}) the value of $\bm\alpha(\bm r_0)$ may thus depend on $\bm\alpha(\bm r)$ at any point on these trajectories. This means that we have no a priori knowledge of the form of the Jacobian of Eq.\ (\ref{eq:num_alpha}) and sophisticated methods of solving $\bm\alpha$ are not available to us. We solve $\bm\alpha$ using a simple fixed-point iteration
\begin{equation}
\bm{\alpha}^{(k+1)} = \bm{\alpha}^{(k)} - \frac{\bm{\alpha}^{(k)} - \bm f(\bm{\alpha}^{(k)}) }{1 - M},
\end{equation}
where $\bm f(\bm\alpha)$ is the right-hand side of Eq.\ (\ref{eq:num_alpha}). This is akin to the Newton-Raphson method \cite{NumericalRecipes}, but instead of calculating the derivative of $\bm f(\bm\alpha)$ we approximate it with the parameter $M\sim 2$. We begin the iteration with $\bm\alpha^{(0)}=\bm v_0$.

We use the finite difference method to solve the continuity equation (\ref{eq:num_poissons}) in cylindrical coordinates. Because of symmetry conditions, we only need to solve the system in a single quadrant of the space surrounding the cylinder, $0\leq\varphi\leq\pi/2$ and $1\leq r\leq R_\infty$. Here $R_\infty$ is a computational cutoff radius. The symmetry conditions translate to boundary conditions $(\partial_\varphi \delta\psi)(r, \varphi=0) = \delta\psi(r, \varphi=\pi/2)=0$. In addition there should be no flow through the surface of the cylinder and the flow should not be modified far from the cylinder. For $\delta\psi$ these conditions mean $(\partial_r \delta\psi)(r=1, \varphi)=(\partial_r \delta\psi)(r=R_\infty, \varphi)=0$. The cutoff radius $R_\infty$ needs to be sufficiently large in order to satisfy the latter condition. We use $R_\infty = 6$ and our lattice spacings are $\delta r \sim 10^{-2}$ and $\delta\varphi \sim 10^{-2}$. The lattice is more tightly spaced close to the cylinder surface.

We can expect the magnitude of $\delta\bm v_s$ to approach zero as we move away from the cylinder. For cylinder velocities below $v_L/2$ the term on the right-hand side of Eq.\ (\ref{eq:num_poissons}) disappears. This means that, due to our boundary conditions, the quantity $\delta\psi$ is identically zero and the ideal flow is unmodified. For cylinder velocities above $v_L/2$ but under $v_L$ the Landau velocity is exceeded locally, which leads to deviation from the ideal flow in the near region. For cylinder velocities above $v_L$ the Landau velocity is exceeded even far from the cylinder, but the right-hand side of Eq.\ (\ref{eq:num_poissons}) still approaches zero at large distances as the solid angle covered by the cylinder becomes small.

We employ the method of successive under-relaxation \cite{NumericalRecipes} in an attempt to introduce stability to our iterative process. Potential $\delta\psi$ at iteration step $k$ is
\begin{equation}
\delta\psi^{(k)} = \tau\delta\psi^{(k)}_C  + (1 - \tau)\delta\psi^{(k-1)},
\end{equation}
where $\delta\psi^{(k)}_C$ is solved from the continuity equation (\ref{eq:num_poissons}) using $\bm I$ at iteration step $k$. We start the iteration from $\delta\psi^{(0)} = 0$. For under-relaxation, the relaxation parameter $\tau \in \left] 0,1 \right[$. All numerical results use $F_1^s=5.4$, which is the zero pressure value in liquid $^3$He.

\begin{figure*}[tb] %  figure placement: top, bottom, (not here or page)
\centering
\subfloat[Ideal flow]{\label{fig:hydro8} \includegraphics[height=0.3\linewidth]{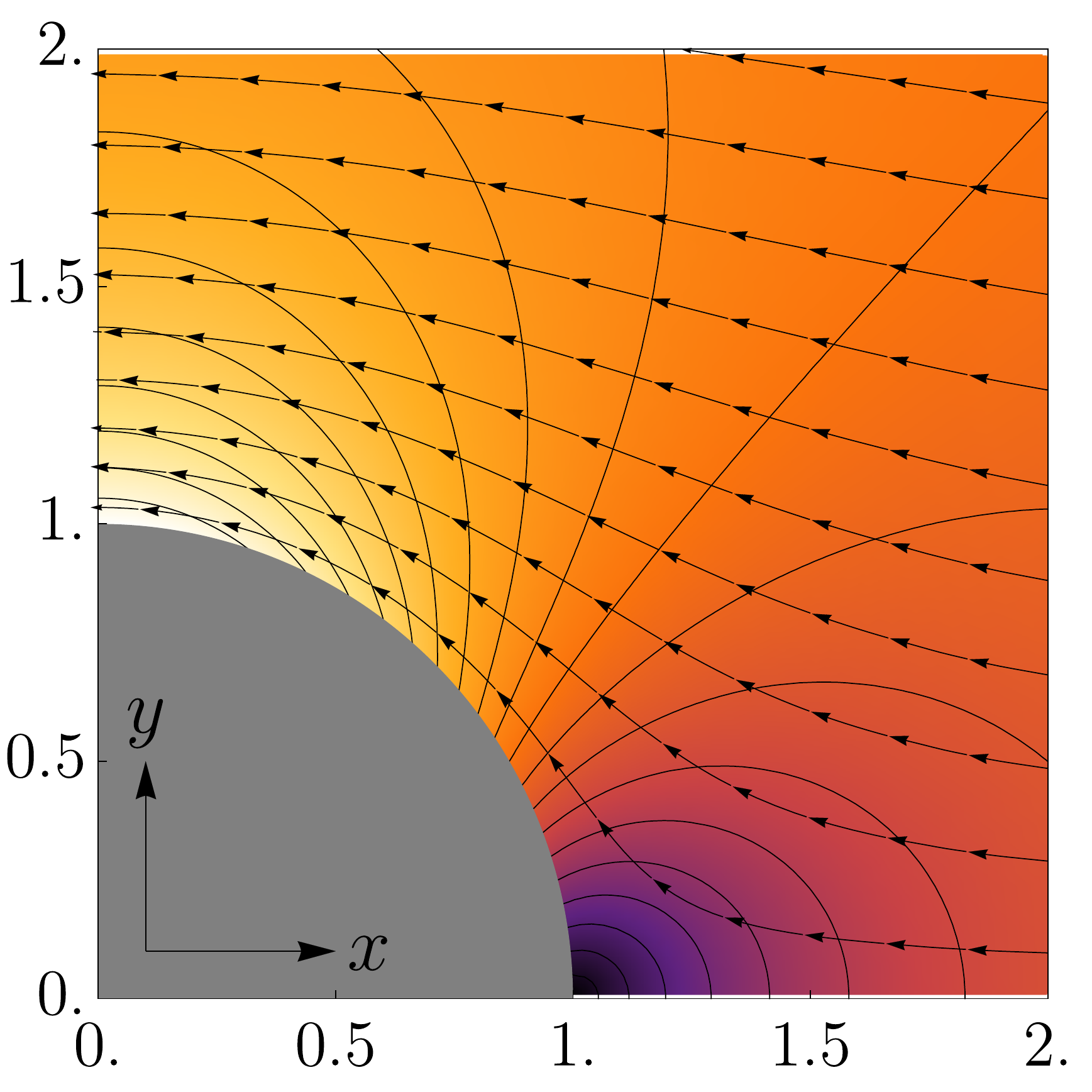}}
\subfloat[Model 1]{\label{fig:model8} \includegraphics[height=0.3\linewidth]{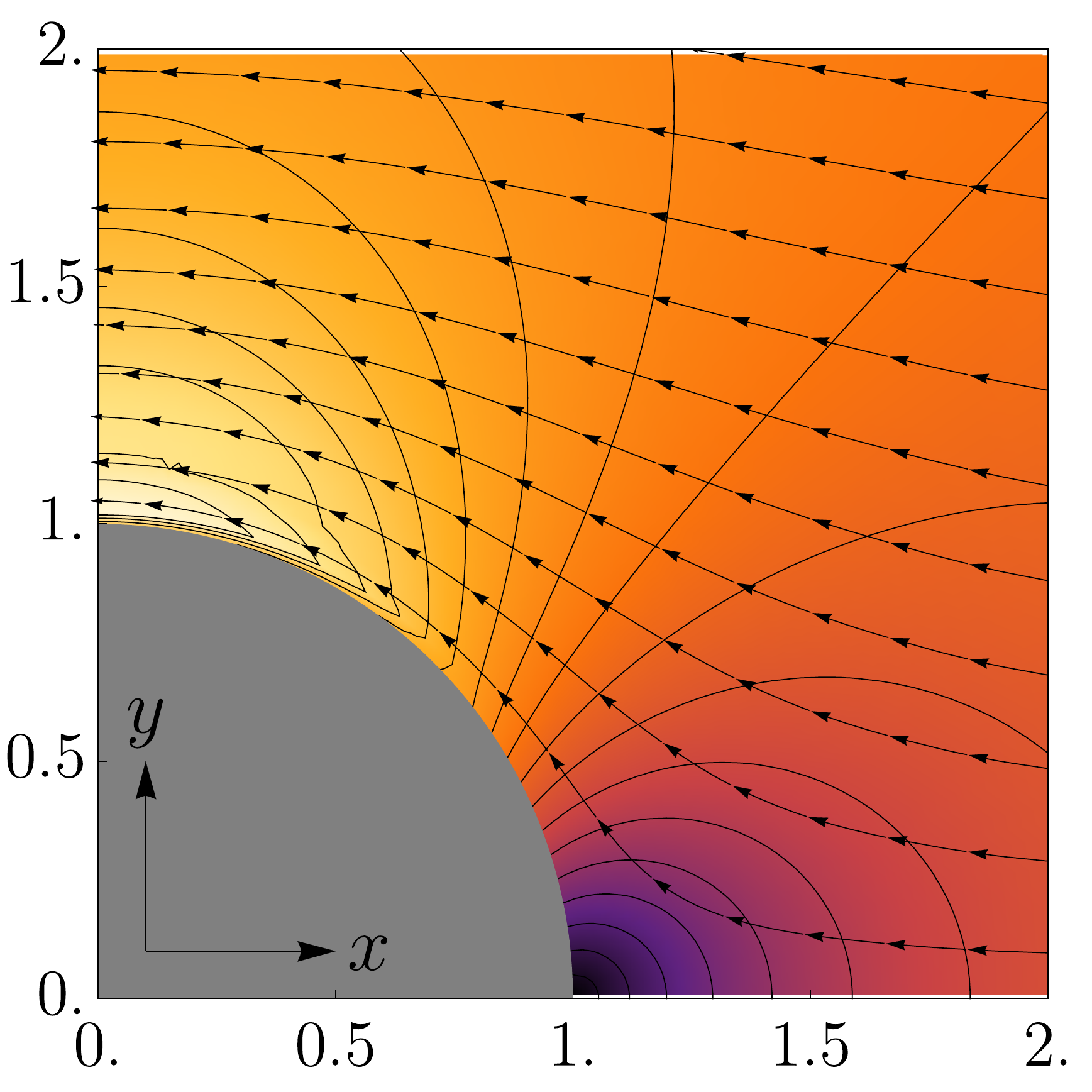}}
\subfloat[Model 2]{\label{fig:yam8} \includegraphics[height=0.3\linewidth]{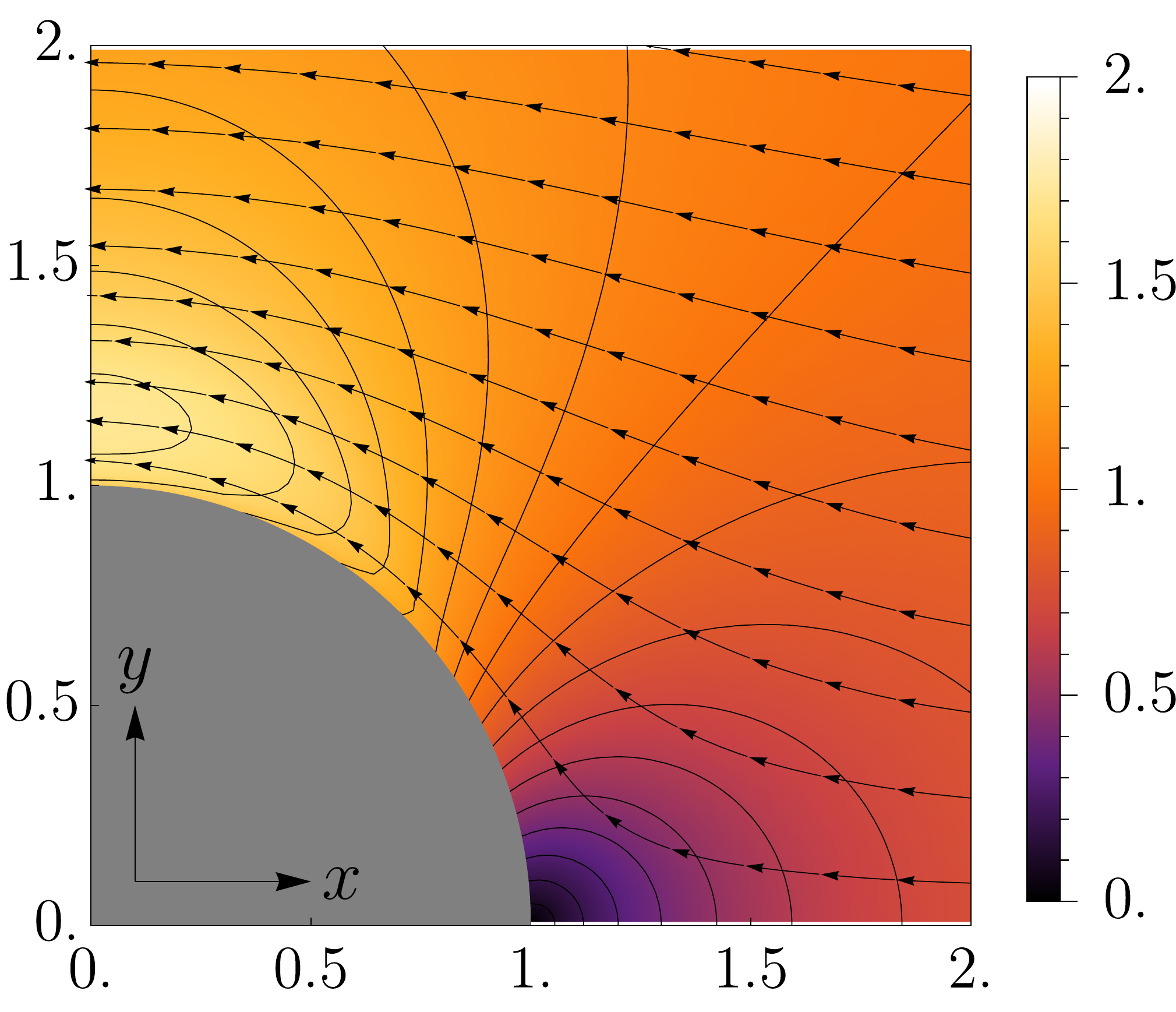}}
\caption{The field $\bm\alpha$ in units of $p_F v$ calculated in three different cases: (a) Ideal-fluid flow field, valid at $v<0.5v_L$, where $\bm \alpha=p_F\bm v_s$ and $\bm v_s$ is given by Eqs.\ (\ref{e.vsgp}) and (\ref{e.iflff}). (b), (c) Two models for self-consistent flow at $v=0.8v_L$. The field $\bm\alpha$ gives the quasiparticle potential $a =\bm\alpha\cdot\hat{\bm p}$ in the dispersion relation (\ref{e.edrel}). The gray segment of a disk represents the cylinder. The color gradient signifies the magnitude of the field, and the stream lines its direction. Contour lines of constant magnitude are also displayed at intervals of $0.1 p_Fv$. The distinguishing feature of the self-consistent models  is the suppression of $\bm{\alpha}$ near the top surface of the cylinder, where locally $v_s>v_L$, to the degree that the field maximum has detached from the cylinder surface. This happens for both models, but is more pronounced in Model 2. In Model 1 this suppression leads to a sharp gradient of $\bm\alpha$ near the cylinder surface.}
\label{fig:alphafield}
\end{figure*}

Results for the self-consistent $\bm\alpha$ using $\bm I_1$ (\ref{eq:collisionless_integral_model_1}) and $\bm I_2$ (\ref{eq:collisionless_integral_model_2}) are shown in Fig.\ \ref{fig:alphafield} together with the ideal flow, for which $\bm\alpha = p_F \bm v_0$. In Model 1 and Model 2, when compared to the ideal flow, the magnitude of $\bm\alpha$ is reduced near the cylinder surface. Excitations in this region cause a nonzero $\bm I$ and as a result $\bm\alpha$ must change in order for the self-consistency equation (\ref{eq:num_alpha}) to be satisfied. Far away from the cylinder the flow is unmodified.

The difference between Model 1 and Model 2 arises from the occupation of K$'$-type states. These do not directly interact with the cylinder surface, but when occupied will locally modify the superflow in a manner that reduces $\bm\alpha$ in the vicinity of the cylinder surface. 

\begin{figure*}[tb] %  figure placement: top, bottom, (not here or page)
\centering
\subfloat[Model 1]{\label{fig:VSmodel8} \includegraphics[height=0.3\linewidth]{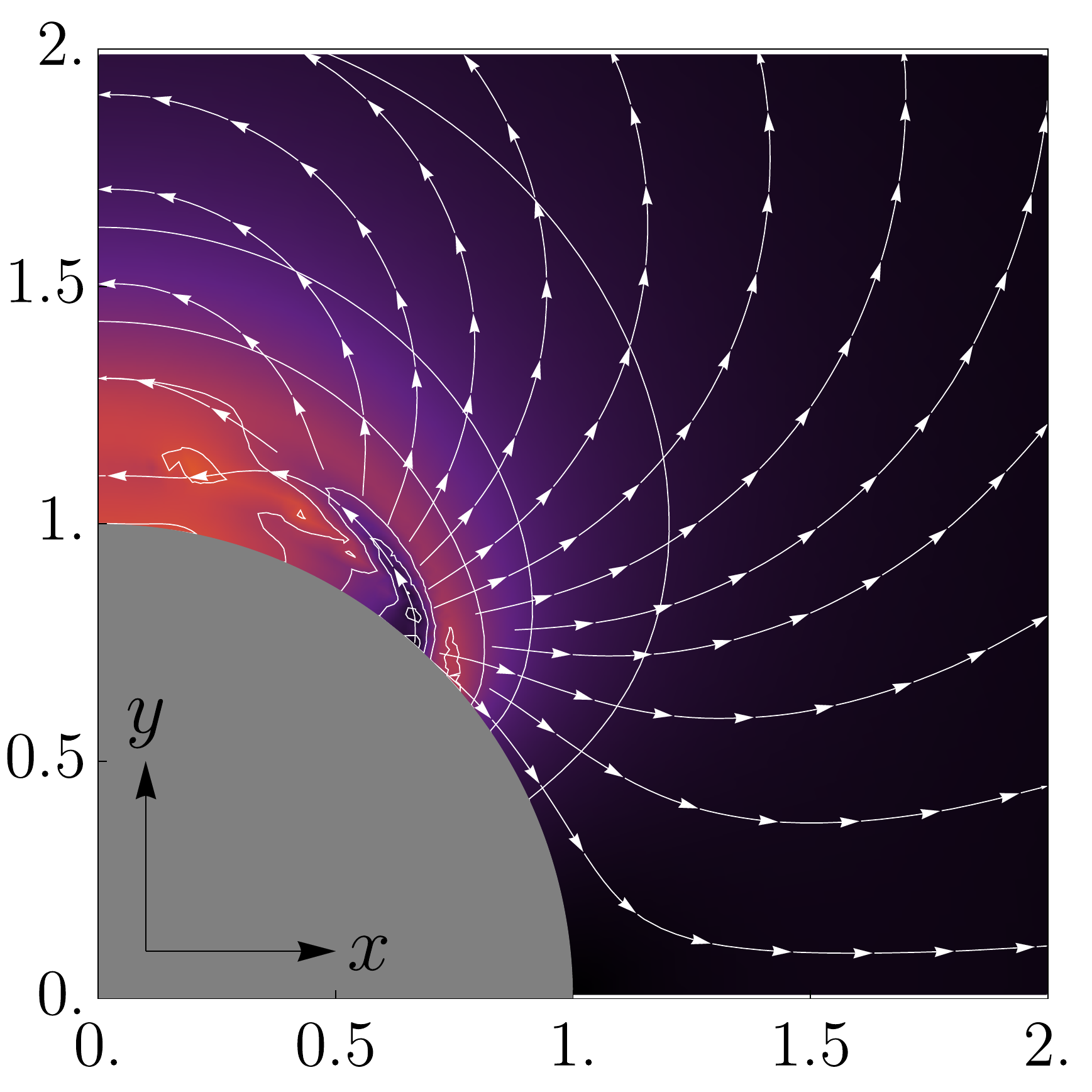}}
\subfloat[Model 2]{\label{fig:VSyam8} \includegraphics[height=0.3\linewidth]{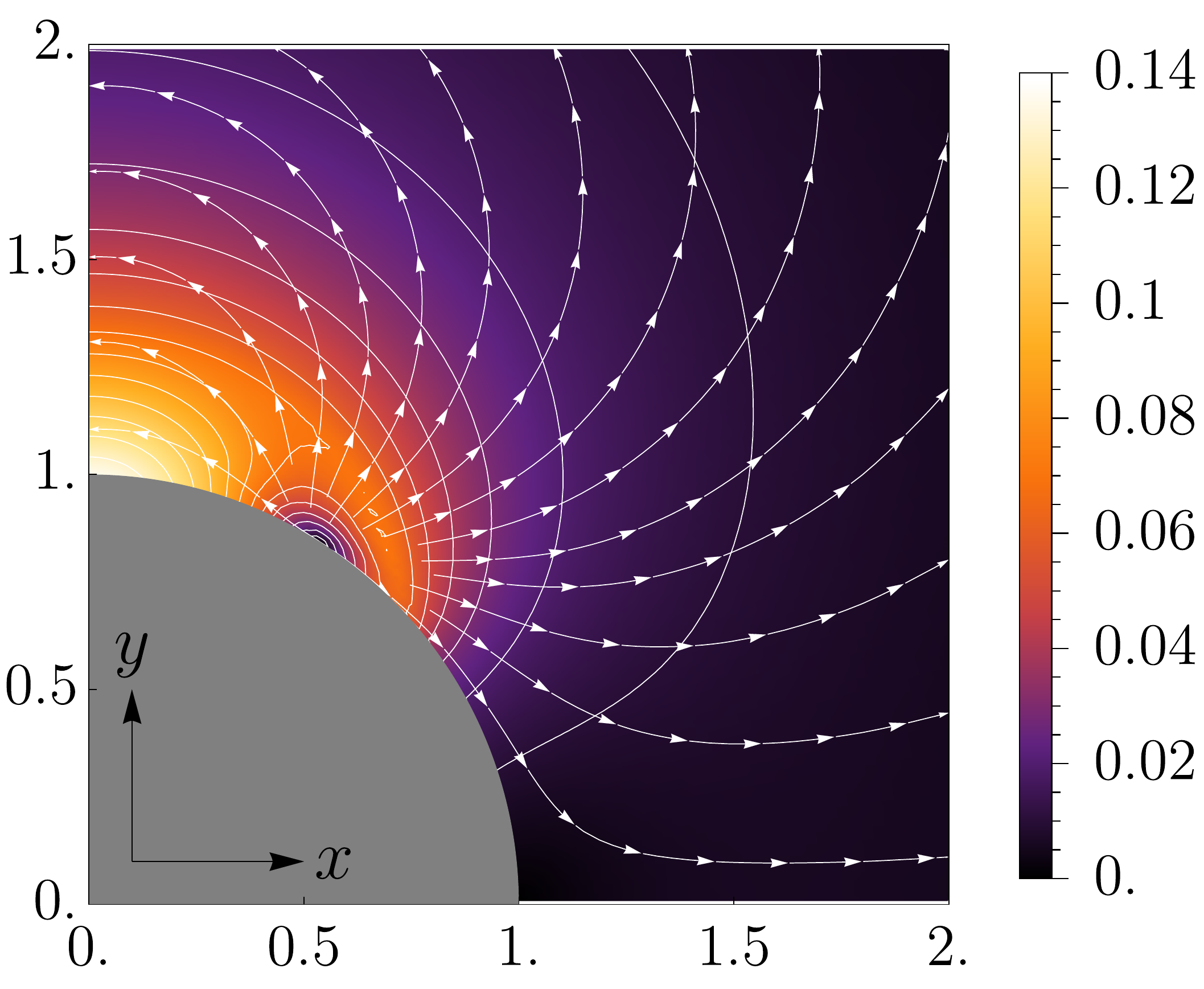}}
\caption{Modification to the ideal flow, $\delta\bm v_s$, in units of $v$ at $v=0.8v_L$, corresponding to the quasiparticle potentials in Fig.\ \ref{fig:alphafield}(b) and Fig.\ \ref{fig:alphafield}(c). The gray segment of a disk represents the cylinder. The color gradient signifies the magnitude of the field in question, and the stream lines its direction. Contour lines of constant magnitude are also displayed at intervals of $0.01 v$. The ideal flow is modified in a manner that diverts the liquid from the regions where Landau velocity is exceeded locally. In addition to the mass flow proportional to $\bm v_s$, there is a quasiparticle current $\propto\bm I$ that on top of the cylinder is to the right. The total current given by Eq.\ (\ref{eq:collisionless_current}) is conserved.}
\label{fig:vsfield}
\end{figure*}

Figure \ref{fig:vsfield} shows $\delta\bm v_s$-fields corresponding to the $\bm\alpha$-fields in Fig.\ \ref{fig:alphafield}(b) and Fig.\ \ref{fig:alphafield}(c). The fields $\delta\bm v_s$ are such that part of the flow is driven to circumvent the areas close to the cylinder surface where Landau velocity is exceeded locally. This reduces the superfluid flow velocity towards the cylinder in the front region and thus increases $\epsilon_3$ (Fig.\ \ref{f.trajectoryenergy}). The same effect is present at higher velocities where it leads to a reduction in the drag force compared to the ideal-fluid case where $\bm I=0$. This effect is greater in Model 2 where K$'$-type states are in equilibrium with the cylinder. Total mass current is given by Eq.\ (\ref{eq:collisionless_current}). The term proportional to $\bm I$ represents a quasiparticle current flowing in a direction opposite to $\bm v_s$, ensuring that mass current is conserved.

After we have solved the self-consistent flow field, we can calculate the force exerted on the cylinder. This is given by Eqs.\ (\ref{eq:collisionless_force}) and (\ref{eq:collisionless_stress_tensor}). Since we have assumed that the scattered distributions are $1/2 - \theta(\epsilon)$ when calculating the flow field, we shall make the same assumption when calculating the force. This means that we can substitute $g A / 2 = 0$ and $g B / 2 = 1/2 - \theta(\epsilon)$ into Eq.\ (\ref{eq:collisionless_stress_tensor}). As we saw in Secs.\ \ref{s.ion} and \ref{s.hydrof}, and in Fig.\ \ref{fig:IonHydroForce}, this will likely underestimate the force compared to using the exact boundary condition, but the difference should be small in the vicinity of the critical velocity.

\begin{figure}[tb] %  figure placement: top, bottom, (not here or page)
\centering
\includegraphics[width=0.8\linewidth]{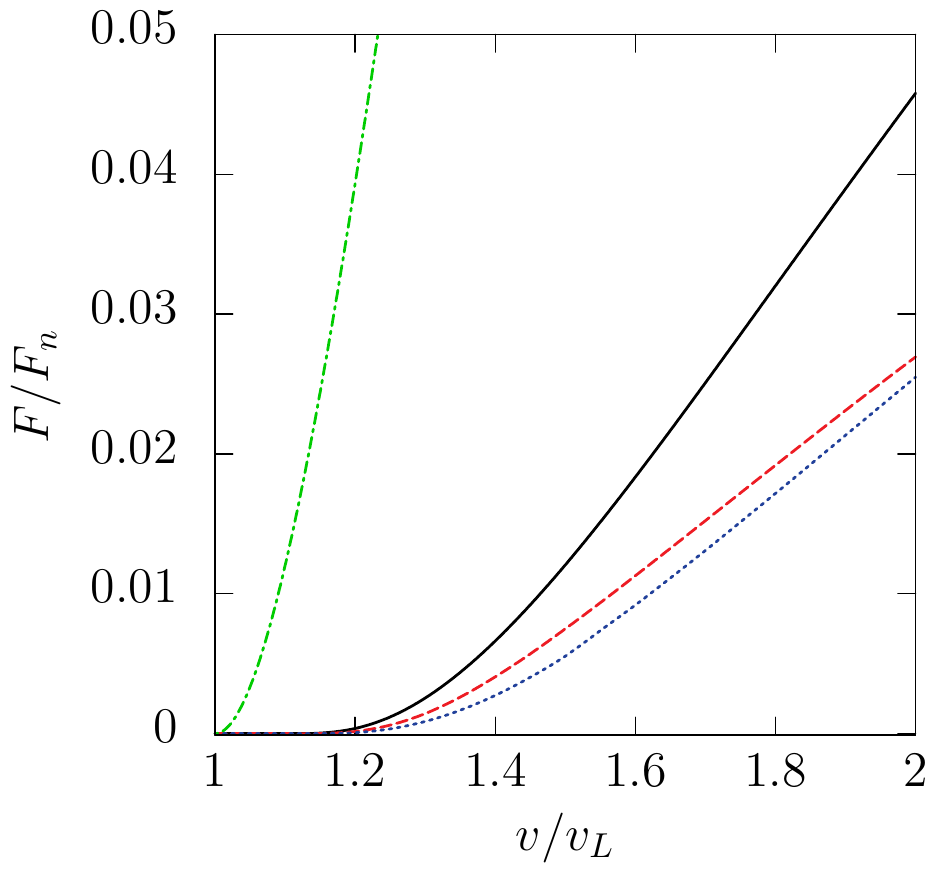} 
\caption{
Drag force exerted on a macroscopic cylinder (diameter $\gg \xi_0$) as a function of velocity in the collisionless approximation. The force is calculated for three different flow fields, the ideal flow (solid black line), Model 1 (dashed red line), and Model 2 (dotted blue line), see Secs.\ \ref{s.hydrof} and \ref{s.noc}. As a comparison, the figure also shows the force exerted on a small pointlike object (diameter $\ll \xi_0$) (dot-dashed green line), see Sec.\ \ref{s.ion}. In all cases we have assumed that the scattered distributions are equilibrium distributions in the object frame, $1 / 2 - \theta(\epsilon)$, since this was used to calculate the flow fields in Model 1 and Model 2. We have also assumed zero temperature, constant gap and $F_1^s = 5.4$, corresponding to zero pressure in liquid $^3$He. 
}
\label{fig:Force}
\end{figure}

Figure \ref{fig:Force} shows the force calculated for the ideal flow field, Model 1, and Model 2. The critical velocities are equal in all three cases, approximately $1.12 v_L$. The self-consistent flow fields, Model 1 and Model 2, both yield a force that is smaller than in the case of the ideal flow field. The force for Model 1 is slightly larger than for Model 2, but the difference between the two is small. The similarity in forces between the two models is likely due to the fact that, despite the differences in the $\bm\alpha$ fields close to the cylinder, the energy $\epsilon_3$ of Fig.\ \ref{f.trajectoryenergy} along quasiparticle trajectories is mostly the same. The states bound to the vicinity of the cylinder affect the force only by means of raising or lowering this energy barrier, which in this case is minimal.

\section{Equilibrium in the near region}\label{s.enr}

In the previous section we studied how the flow field around a cylinder is modified when there are no collisions between quasiparticles. In this section we assume the opposite extreme where, due to quasiparticle-quasiparticle collisions, full equilibrium with the cylinder has been achieved in the near region.
We assume that the far region is still in equilibrium with the laboratory frame, and that the gap amplitude there is isotropic, $|\Delta(\hat{\bm p})| = \Delta$. 
This situation can only be achieved at cylinder velocities $v$ less than the Landau velocity $v_L = \Delta / p_F$.
Unlike in the collisionless case, we allow the gap amplitude to become anisotropic in the near region.
Parametrizing the momentum direction as $\hat{\bm p} = \bm p_\perp + p_\parallel \hat{\bm \alpha}$, where $\bm p_\perp \cdot \hat{\bm \alpha} = 0$, the square of the gap amplitude in a p-wave superfluid can be written as $|\Delta(\hat{\bm p}, \bm r)|^2 = \Delta_\perp(\bm r)^2 p_\perp^2 + \Delta_\parallel(\bm r)^2 p_\parallel^2$.

We shall study the system in the rest frame of the cylinder at $T=0$. The flow obeys the same equations (\ref{eq:collisionless_alpha_eq}), (\ref{eq:collisionless_psi_eq}), and (\ref{eq:collisionless_vs}) as it did in the collisionless case. When calculating $\bm I$, we can use distributions $\phi_{Bi} = 1/2 - \theta(\epsilon)$ everywhere. These describe equilibrium with the cylinder, which is required in the near region. In the far region $\bm I$ vanishes as it should, since there are no excitations present.

Substituting the distributions and the gap amplitude into Eq.\ (\ref{eq:collisionless_integral}) yields
\begin{equation}\label{eq:equilibrium_I}
\bm I = - \frac{1}{3}  \theta(\alpha - \Delta_\parallel) \frac{(\alpha^2 - \Delta_\parallel^2)^{3/2}}{\alpha^2 + \Delta_\perp^2 - \Delta_\parallel^2} \hat{\bm \alpha}.
\end{equation}
We see from Eq.\ (\ref{eq:collisionless_alpha_eq}) that $\bm I$ and $\bm v_s$ are parallel and subsequently we can express the mass current density as
\begin{equation}
\bm j = \rho_s(v_s ) \bm v_s. \label{eq:equilibrium_current}
\end{equation}
Here $\rho_s$ is the superfluid density, which depends on the magnitude of the superfluid velocity. The value of $\rho_s$ is given by
\begin{equation}
\rho_s(v_s) = m n_f - \frac{m n_f}{p_F}(3 + F_1^s) \frac{I\bm(\alpha(v_s)\bm)}{v_s},
\end{equation}
where $\alpha(v_s)$ is the solution of the nonlinear equation
\begin{equation}
\alpha = p_F v_s  - \frac{F_1^s}{3}  \theta(\alpha - \Delta_\parallel) \frac{(\alpha^2 - \Delta_\parallel^2)^{3/2}}{\alpha^2 + \Delta_\perp^2 - \Delta_\parallel^2}.
\end{equation}
We can write Eqs.\ (\ref{eq:collisionless_alpha_eq}),  (\ref{eq:collisionless_psi_eq}), and (\ref{eq:collisionless_vs}) as
\begin{align}
\bm\nabla \cdot \left[\rho_s(v_s ) \bm v_s \right]  &= 0, \label{eq:equilibrium_continuity_eq} \\ 
\bm v_s &= \frac{\hbar}{2 m} \bm\nabla \psi. \label{eq:equilibrium_velocity}
\end{align}

\begin{figure}[tb] %  figure placement: top, bottom, (not here or page)
\centering
\includegraphics[width=0.8\linewidth]{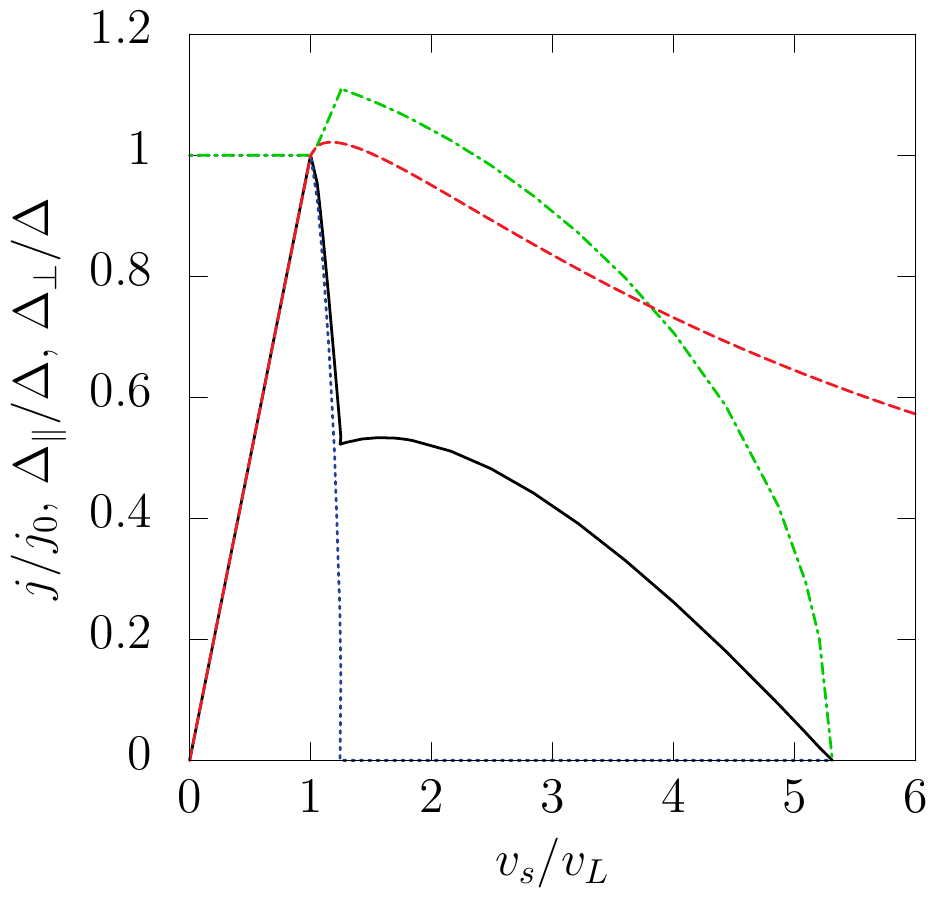} 
\caption{The equilibrium mass current $j$ in units of $j_0=m n_f v_L$ for self-consistent gap (solid black line) and for constant gap $\Delta$ (dashed red line), together with the parallel and perpendicular components of the self-consistent gap, $\Delta_\parallel$ (dotted blue line) and $\Delta_\perp$ (dot-dashed green line), as functions of superfluid velocity $v_s$, with $F_1^s = 5.4$. See Refs.\ \cite{Vollhardt80,Kleinert80} for details.}
\label{fig:jDelta}
\end{figure}

So far we have not specified the gap functions $\Delta_\parallel$ and $\Delta_\perp$. Obviously the correct choice would be to determine $\Delta_\parallel$ and $\Delta_\perp$ self-consistently from the gap equation. This has been done in  Refs.\ \cite{Vollhardt80}, \cite{Kleinert80}. The result is that for $F_1^s = 5.4$ both
$\Delta_\parallel$ and $\Delta_\perp$, as well as $j$, are single-valued functions of $v_s$, see Fig.\ \ref{fig:jDelta}. (This is not the case for $F_1^s = 0$.) For $v_s<v_L$ the gap components are constants and $j$ grows linearly. Increasing $v_s$ beyond $v_L$, the parallel gap component $\Delta_\parallel$ drops rapidly to zero. The perpendicular gap component $\Delta_\perp$ first grows slightly. Current $j$ drops sharply, recovering minutely as $\Delta_\perp$ begins to decrease. Both $\Delta_\perp$ and $j$ go to zero at $v_s=5.3v_L$.

The equilibrium problem now consists of finding the solution of Eqs.\ (\ref{eq:equilibrium_continuity_eq}) and  (\ref{eq:equilibrium_velocity}) with proper boundary conditions. At $v<v_L/2$ the solution is the ideal superfluid flow (\ref{e.iflff}). We have attempted to find a solution numerically at $v>v_L/2$, but have been unsuccessful.

In order to gain insight into the failure to find a stable solution at $v>v_L/2$, we have considered an alternative model. Instead of using the gap equation, we have assumed a constant gap $\Delta_\parallel = \Delta_\perp = \Delta$, which is independent of $v_s$. As seen in Fig.\ \ref{fig:jDelta}, in this case the mass current $j$ first increases slightly beyond $v_L$, but after that is a monotonically decreasing function of $v_s$. It drops more slowly than in the case of the self-consistent gap and does not vanish at any finite $v_s$.

\begin{figure*}[tb] %  figure placement: top, bottom, (not here or page)
%\centering
\subfloat[$\bm\alpha$]{\label{fig:equilibrium65} \includegraphics[height=0.3\linewidth]{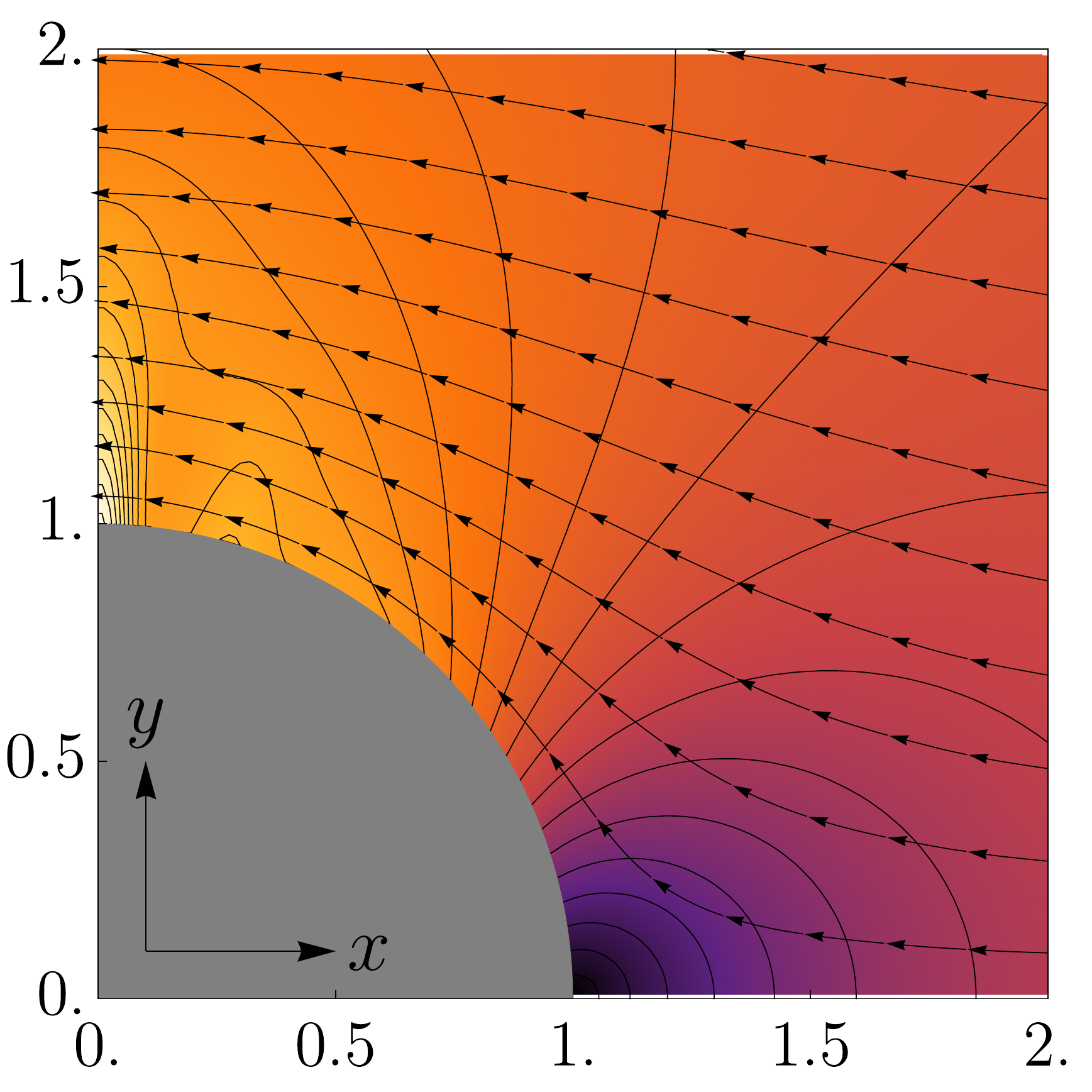}}
\subfloat[$\delta\bm v_s$]{\label{fig:VSequilibrium65} \includegraphics[height=0.3\linewidth]{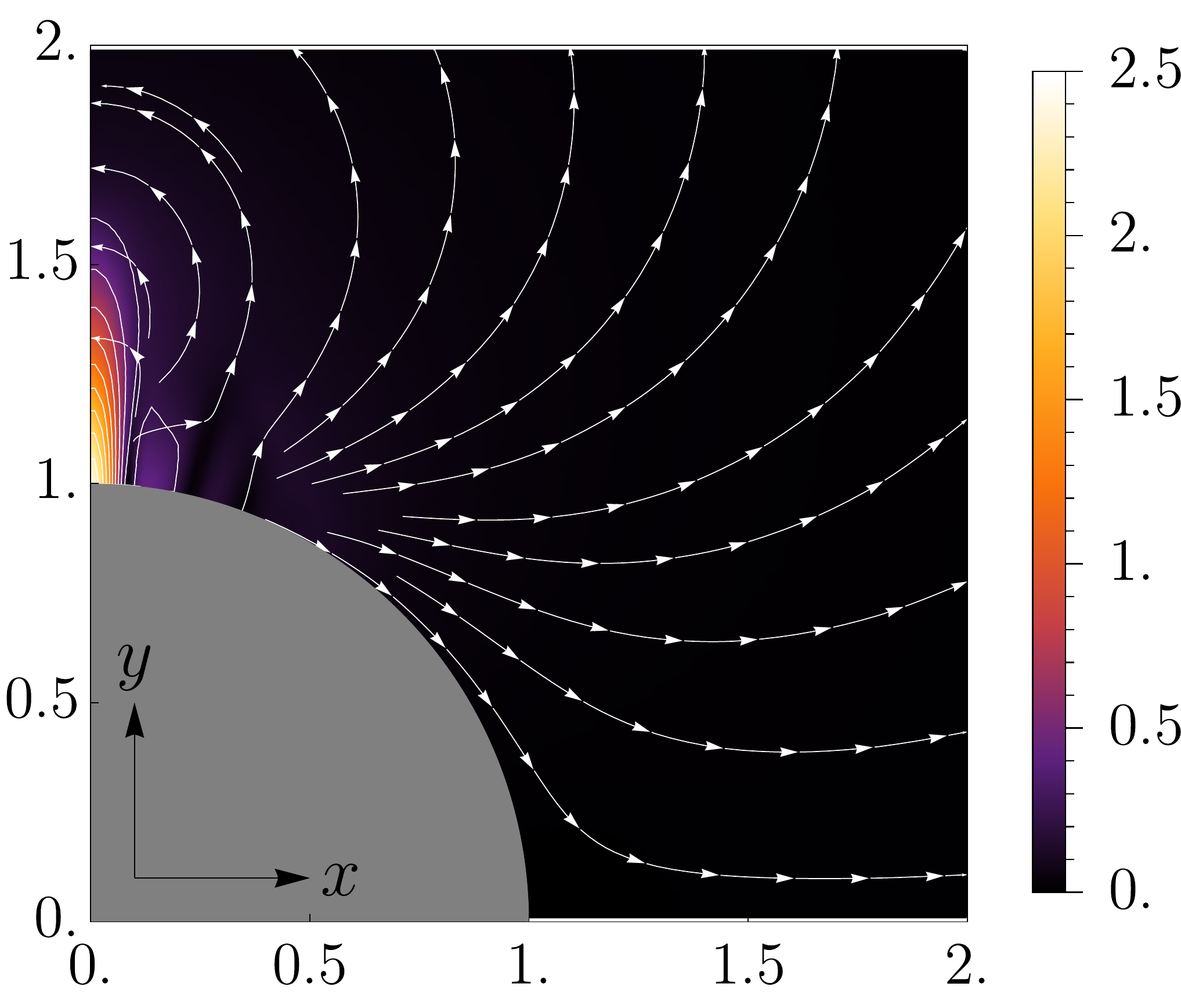}}
\caption{Equilibrium in the near region assuming constant $\Delta$: (a) $\bm\alpha$ in units of $p_F v$ and (b) $\delta\bm v_s$ in units of $v$. The parameters are $v = 0.65 v_L$, $T=0$, and $F_1^s=5.4$. For $\bm\alpha$, the contour lines are placed at intervals of $0.1 p_F v$ and for $\delta\bm v_s$ at intervals of $0.25 v$. Note that the magnitude of $\delta\bm v_s$ is much greater than in the collisionless case (Fig.\ \ref{fig:vsfield}), in spite of the smaller cylinder velocity.}\label{fig:equilibriumfields}
\end{figure*}

Figure \ref{fig:equilibriumfields} displays numerical results for fields $\bm\alpha$ and $\delta\bm v_s = \bm v_s - \bm v_0$ in the case of the constant gap at $v = 0.65 v_L$. We see that, unlike in the collisionless approximation, there is a region near the top surface of the cylinder where the magnitude of $\bm \alpha$ is larger than in the case of the ideal flow. 
In the same region the length of $\delta \bm v_s$ is an order of magnitude larger than in the collisionless approximation. In fact, $\delta v_s$ is of the same order as $v_0$, thus significantly changing the flow pattern. Outside this region the flow is not significantly modified.
For our preferred lattice the calculation becomes nonconvergent at velocities slightly greater than $0.7 v_L$. Decreasing the lattice spacing leads to nonconvergence at even lower velocities. 

What causes the difference between the collisionless models and the equilibrium models is the number of excitations present in the flow field, which can be much larger in the latter case. This also explains the instability of the flow in the equilibrium models.
In order to satisfy conservation of mass, the local mass current on the $y$-axis has to be larger than the far-region value $j_\infty = m n_f v$. In ideal fluid $\bm j \propto \bm v_s$, and thus the mass current can be increased simply by increasing the superfluid velocity. Increasing $v_s$ beyond the Landau velocity also builds up the excitation current $\propto \bm I$ opposite to the direction of superflow, decreasing the total mass flow. In the collisionless approximation $I$ is generally small, since only those trajectories that collide with the object contribute to it. This means that increasing $v_s$ still increases $j$, although at a smaller rate than for ideal fluid. In the equilibrium model, however, $I$ is so large that $j$ starts to decrease when $v_s$ is increased from $\sim v_L$, meaning that $j$ has a maximum value. Because of this, the equilibrium models have no solution at high enough velocities, since it is not possible to satisfy conservation of mass. The difference in convergence between the two equilibrium models seems to have the same origin. In addition to these two, we have tested other $j(v_s)$ functions. It seems that especially the sharp drop in $j(v_s)$ at $v_s=v_L$ is the cause of the instability of the self-consistent-gap model.

The lack of convergence apparently means that the true physical solution is not consistent with the assumptions made. The theory presented here is limited to long length scale, time-independent solutions with a singly defined phase $\psi$. Thus there could be an instability to some time-dependent state that could have vortex-type structures.

\section{Summary}

We have investigated objects moving in superfluid Fermi liquid at velocities on the order of the Landau velocity. The prevailing assumption that an object exceeding the Landau velocity would experience a sudden onset of drag force seems to hold true only for objects  much smaller than the coherence length. For a large object the fluid has to be pushed away from the object's path, resulting in a spatially varying flow field that affects the quasiparticle energies. This leads to Andreev reflections that prevent some excitations from escaping into the surrounding fluid. Perhaps counterintuitively, the critical velocity is increased and the drag force decreased. 

This work was to a large extend motivated by the experiment by Bradley et al.\ \cite{Bradley16}. The drag force they measure is reduced from the normal-state value by a factor on the order of $10^{-5}$ at $v=2v_L$. In the absence of collisions between quasiparticles, we calculate theoretically a reduction factor on the order of  $10^{-2}$. In the opposite, collision-dominated limit our calculation implies an instability. The main theoretical problem concerning the interpretation of the experiment is whether there could be an intermediate state between the two limiting cases, one with additional shielding factor three orders of magnitude greater compared to the collisionless limit. Also, what is the nature of this state? Does it contain vortices and in what configuration? Is it a stable or a transient state?

There are several limitations in the present work. It is limited to time-independent states. Only two extreme cases of quasiparticle-quasiparticle collisions were studied. Vortex-like structures were excluded. We have used a simple boundary condition that ignores quantum processes such as Andreev reflection and generation of magnetic excitations in the surface layer. We considered only uniform motion, which leaves the occupation of trapped states ambiguous in the collisionless limit.  The assumption of a constant, isotropic gap was also made in the collisionless limit. We have assumed zero temperature and a steplike distribution for the scattered quasiparticles. The work presented here should be seen as the first theoretical study of the drag force on a macroscopic object exceeding the Landau velocity in a Fermi superfluid. We hope our work stimulates further experimental and theoretical studies on this topic.

\begin{acknowledgments}
We thank S. Autti, V. Eltsov, R. Haley, M. Krusius, Y. Lee, G. Pickett, J. Sauls, G. Volovik and D. Zmeev for useful discussions. This work was financially supported by the Vilho, Yrj\"o and Kalle V\"ais\"al\"a Foundation, the Jenny and Antti Wihuri Foundation and the Oskar \"Oflunds Stiftelse sr. 
\end{acknowledgments}


\begin{thebibliography}{99}

\bibitem{Mach1887} E. Mach and P. Salcher, {\it Photographische Fixirung der durch Projectile in der Luft eingeleiteten Vorg\"ange}, Sitzungsber. Kaiserl. Akad. Wiss. Wien, Math.-Naturwiss. Cl. {\bf 95} Abt. II, 764 (1887).

\bibitem{LL:EDCM} L. D. Landau and E. M. Lifshitz, {\it Electrodynamics of Continuous Media} (Pergamon Press, Oxford, 1984).

\bibitem{Thomson1891} W. Thomson, {\it Popular Lectures and Addresses} (MacMillan, London, 1891), Vol. III, pp. 450-500.

\bibitem{Whitham} G. B. Whitham, {\it Linear and Nonlinear Waves} (Wiley, New York, 1974).

\bibitem{Landau41} L. Landau, J. Phys. USSR {\bf 5}, {\it The Theory of Superfluidity of Helium II}, 71 (1941).

\bibitem{Allum77} D. R. Allum, P. V. E. McClintock, A. Phillips, and R. M. Bowley, {\it The Breakdown of Superfluidity in Liquid {\rm $^4$He}: an Experimental Test of Landau's Theory}, Phil. Trans. R. Soc. London A {\bf 284}, 179 (1977).

\bibitem{Ahonen76} A. I. Ahonen, J. Kokko, O. V. Lounasmaa, M. A. Paalanen, R. C. Richardson, W. Schoepe, and Y. Takano, {\it Mobility of Negative Ions in Superfluid {\rm $^3$He}}, Phys. Rev. Lett. {\bf 37}, 511 (1976).

\bibitem{Donnelly91} R. J. Donnelly, {\it Quantized Vortices in Helium II} (Cambridge, 1991).

\bibitem{Castelijns86}  C. A. M. Castelijns, K. F. Coates, A. M. Gu\'enault, S. G. Mussett, and G. R. Pickett, {\it Landau Critical Velocity for a Macroscopic Object Moving in Superfluid {\rm $^3$He}-B: Evidence for Gap Suppression at a Moving Surface}, Phys. Rev. Lett. {\bf 56}, 69 (1986).

\bibitem{Ruutu97} V. M. H. Ruutu, \"U. Parts, J. H. Koivuniemi, N. B. Kopnin, and M. Krusius, {\it Intrinsic and Extrinsic Mechanisms of Vortex Formation in Superfluid {\rm $^3$He}-B}, J. Low Temp. Phys. {\bf 107}, 93 (1997).

\bibitem{Onofrio00} R. Onofrio, C. Raman, J. M. Vogels, J. R. Abo-Shaeer, A. P. Chikkatur, and W. Ketterle, {\it Observation of Superfluid Flow in a Bose-Einstein Condensed Gas}, Phys. Rev. Lett. {\bf 85}, 2228 (2000).

\bibitem{Varoquaux15} E. Varoquaux, {\it Anderson's Considerations on the Flow of Superfluid Helium: Some Offshoots}, Rev. Mod. Phys. {\bf 87} 803 (2015)

\bibitem{Bradley16} D. I. Bradley, S. N. Fisher, A. M. Gu\'enault, R. P. Haley, C. R. Lawson, G. R. Pickett, R. Schanen, M. Skyba, V. Tsepelin, and D. E. Zmeev, {\it Breaking the Superfluid Speed Limit in a Fermionic Condensate}, Nat. Phys. {\bf 12}, 1017 (2016).

\bibitem{LLf} L. D. Landau and E. M. Lifshitz, {\it Fluid Mechanics} (Pergamon Press, Oxford, 1987).

\bibitem{Batchelor} G. K. Batchelor, {\it An Introduction to Fluid Dynamics} (Cambridge, 1967).

\bibitem{Serene83} J. W. Serene and D. Rainer, {\it The Quasiclassical Approach to Superfluid {\rm $^3$He}}, Phys. Rep. {\bf 101}, 221 (1983).

\bibitem{Combescot06} R. Combescot, M. Yu. Kagan, and S. Stringari, {\it Collective Mode of Homogeneous Superfluid Fermi Gases in the BEC-BCS Crossover}, Phys. Rev. A {\bf 74}, 042717  (2006). 

\bibitem{Tinkham} M. Tinkham, {\it Introduction to Superconductivity} (McGraw-Hill, New York, 1996), Second Edition.

\bibitem{Bardeen62} J. Bardeen, {\it Critical Fields and Currents in Superconductors}, Rev. Mod. Phys. {\bf 34}, 667 (1962).

\bibitem{Zagoskin} A. M. Zagoskin, {\it Quantum Theory of Many-Body Systems: Techniques and Applications}
(Springer, New York, 1998).

\bibitem{Vollhardt80} D. Vollhardt, K. Maki, and N. Schopohl, {\it Anisotropic Gap Distortion Due to Superflow and the Depairing Critical Current in Superfluid {\rm $^3$He}-B}, J. Low Temp. Phys. {\bf 39}, 79 (1980).

\bibitem{Kleinert80} H. Kleinert, {\it Depairing Critical Current of {\rm $^3$He}-B at All Temperatures Including Gap Distortion}, J. Low Temp. Phys. {\bf 39}, 451 (1980).

\bibitem{Baym12} G. Baym and C. J. Pethick, {\it Landau Critical Velocity in Weakly Interacting Bose Gases}, Phys Rev. A {\bf 86}, 023602 (2012).

\bibitem{Leggett75} A.J. Leggett, {\it A Theoretical Description of the New Phases of Liquid {\rm $^3$He}}, Rev. Mod. Phys. {\bf 47}, 331 (1975).

\bibitem{Lambert90} C. Lambert, {\it On the Approach to Criticality of a Vibrating, Macroscopic Object in Superfluid {\rm $^3$He}-B}, Physica B {\bf 165}\&{\bf 166}, 653
(1990).

\bibitem{Lambert92} C. Lambert, {\it Theory of Pair Breaking by Vibrating Macroscopic Objects in Superfluid {\rm $^3$He}}, Physica B {\bf 178}, 294 (1992).

\bibitem{Volovik09} G. Volovik, Pis'ma Zh. Eksp. Teor. Fiz. {\bf 90}, {\it Fermion Zero Modes at the Boundary of Superfluid {\rm $^3$He}-B}, 440 (2009) [JETP Lett. {\bf 90}, 398 (2009)].

\bibitem{Zheng17} P. Zheng, W. G. Jiang, C. S. Barquist, Y. Lee, and H. B. Chan, {\it Critical Velocity in the Presence of Surface Bound States in Superfluid {\rm $^3$He}-B}, Phys. Rev. Lett. {\bf 118}, 065301 (2017).

\bibitem{Ashauer88} B. Ashauer, {\it Branch Imbalance Caused by Moving Ions in Superfluid {\rm $^3$He}-B}, J. Phys. C {\bf 21}, 5129 (1988).

\bibitem{Andreev64} A. F. Andreev, {\it The Thermal Conductivity of the Intermediate State in Superconductors}, Zh. Eksp. Teor. Fiz. {\bf 46}, 1823 (1964) [Sov. Phys. JETP {\bf 19}, 1228 (1964)].

\bibitem{Fisher89}  S. N. Fisher, A. M. Gu\'enault, C. J. Kennedy, and G. R. Pickett, {\it Beyond the Two-Fluid Model: Transition from Linear Behavior to a Velocity-Independent Force on a Moving Object in {\rm $^3$He}-B}, Phys. Rev. Lett. {\bf 63}, 2566 (1989).

\bibitem{Fisher91}  S. N. Fisher, G. R. Pickett, and R. J. Watts-tobin, {\it A Microscopic Calculation of the Force on a Wire Moving Through Superfluid {\rm $^3$He}-B in the Ballistic Regime}, J. Low Temp. Phys. {\bf 83}, 225 (1991).

\bibitem{Enrico95}  M. P. Enrico, S. N. Fisher, and R. J. Watts-Tobin, {\it Diffuse Scattering Model of the Thermal Damping of a Wire Moving Through Superfluid {\rm $^3$He}-B at Very Low Temperatures}, J. Low Temp. Phys. {\bf 98}, 81 (1995).

\bibitem{Enrico96}  M. P. Enrico and R. J. Watts-Tobin, {\it Specular and Diffuse Scattering of Quasiparticles by a Macroscopic Object Moving through Superfluid {\rm $^3$He}-B}, J. Low Temp. Phys. {\bf 102}, 103 (1996).

\bibitem{Kieselmann83} G. Kieselmann and D. Rainer, {\it Branch Conversion at Surfaces of Superfluid {\rm $^3$He}}, Z. Phys. B {\bf 52}, 267 (1983).

\bibitem{Kurkijarvi90} J. Kurkij\"arvi and D. Rainer, in {\it Helium Three}, edited by W.P. Halperin and L.P. Pitaevskii (Elsevier, Amsterdam, 1990), pp. 313-352. 

\bibitem{Nagai08} K. Nagai, Y. Nagato, M. Yamamoto, and S. Higashitani, {\it Surface Bound States in Superfluid {\rm $^3$He}}, J. Phys. Soc. Jpn. {\bf 77}, 111003 (2008). 

\bibitem{Shelankov00} A. Shelankov and M. Ozana, {\it Quasiclassical Theory of Superconductivity: A Multiple-Interface Geometry}, Phys. Rev. B {\bf 61}, 7077 (2000).

\bibitem{Eschrig09} M. Eschrig, {\it Scattering Problem in Nonequilibrium Quasiclassical Theory of Metals and Superconductors: General Boundary Conditions and Applications}, Phys. Rev. B {\bf 80}, 134511 (2009).

\bibitem{Zhang88} W. Zhang, J. Kurkij\"arvi, D. Rainer, and E. V. Thuneberg, {\it Andreev Scattering at a Rough Surface of {\rm $^3$He}-B}, Phys. Rev. B {\bf 37}, 3336 (1988).

\bibitem{Ashauer89} B. Ashauer, {\it Emission von Quasiteilchen durch bewegte Objekte in suprafluidem $^3$Helium-B}, Ph.D.\ thesis, University of Bayreuth, 1989.

\bibitem{Fetter87} A. L. Fetter, in {\it The Physics of Liquid and Solid Helium, Part 1}, edited by
K. H. Bennemann and J. B. Ketterson (Wiley, New York, 1978), pp. 207-305. 

\bibitem{Borghesani07} A. F. Borghesani, {\em Ions and Electrons in Liquid Helium} (Oxford, 2007).

\bibitem{Baym79} G. Baym, C. J. Pethick, and M. Salomaa, {\it Mobility of Negative Ions in Superfluid {\rm $^3$He}-B}, J. Low Temp. Phys {\bf 36}, 431 (1979).
 
\bibitem{Salomaa80b} M. Salomaa, C. J. Pethick, and G. Baym, {\it Mobility Tensor of Negative Ions in Superfluid {\rm $^3$He}-A}, J. Low Temp. Phys. {\bf 40}, 297 (1980).

\bibitem{TKR81} E. V. Thuneberg, J. Kurkij\"arvi, and D. Rainer, {\it Quasiclassical Theory of Ions in {\rm $^3$He}}, J. Phys. C {\bf 14}, 5615 (1981).

\bibitem{Shevtsov17} O. Shevtsov and J. A. Sauls, {\it Electron Bubbles and Weyl Fermions in Chiral Superfluid {\rm $^3$He}-A}, Phys. Rev. B {\bf 94}, 064511 (2016).

\bibitem{Tsutsumi17} Y. Tsutsumi, {\it Scattering Theory on Surface Majorana Fermions by an Impurity in {\rm $^3$He}-B}, Phys. Rev. Lett. {\bf 118}, 145301 (2017).

\bibitem{Bowley77} R. M. Bowley, {\it Motion of Negative Ions in Superfluid {\rm $^3$He}}, J. Phys. C {\bf 10}, 4033 (1977).

\bibitem{Rainer87} D. Rainer and B. Ashauer, {\it Quasiparticle Beams with Branch Imbalance in Superfluid {\rm $^3$He}}, Jpn. J. Appl. Phys. {\bf 26}-3, 173 (1987).

\bibitem{Virtanen06} T. H. Virtanen and E. V. Thuneberg, {\it Force on a Slow Object in a Fermi Liquid in the Ballistic Limit}, AIP Conf. Proc. {\bf 850}, 113 (2006).

\bibitem{NumericalRecipes} W. H. Press, S. A. Teukolsky, W. T. Vetterling, and B. P. Flannery, {\it Numerical Recipes: The Art of Scientific Computing} (Cambridge, 2007), Third Edition.

\end{thebibliography}
\end{document}